\documentclass[amsmath,amssymb,prd,notitlepage,longbibliography,superscriptaddress,nofootinbib]{revtex4-2}

\usepackage{amsmath}
\usepackage{enumerate}
\usepackage{amsfonts}
\usepackage[utf8]{inputenc}
\usepackage[T1]{fontenc}
\usepackage{hyperref}

\usepackage{bbm}
\usepackage{amssymb}
\usepackage{amsthm}
\usepackage{mathtools}
\usepackage{filecontents}
\usepackage{comment}
\usepackage{mathrsfs}

\makeatletter
\def\@fpheader{\relax}
\makeatother

\newcounter{parentsubequation}

\makeatother

\DeclareMathAlphabet{\mathbbold}{U}{bbold}{m}{n} 

\DeclareMathOperator{\Tr}{Tr}

\begin{document}

\title{Confluent conformal blocks and the Teukolsky master equation}

\author{Bruno Carneiro da Cunha,}
\email{bruno.ccunha@ufpe.br}
\author{João Paulo Cavalcante}
\email{joaopaulocavalcante@hotmail.com.br}
\affiliation{Departamento de F\'{\i}sica, Universidade Federal de
  Pernambuco, 50670-901, Recife, Brazil}

\begin{abstract}
  Quasinormal modes of usual, four dimensional, Kerr black
  holes are described by certain solutions of a confluent Heun
  differential equation. In this work, we express these solutions in
  terms of the connection matrices for a Riemann-Hilbert problem,
  which was recently solved in terms of the Painlevé V
  transcendent. We use this formulation to generate small-frequency
  expansions for the angular spheroidal harmonic eigenvalue, and
  derive conditions on the monodromy properties for the radial modes.
  Using exponentiation, we relate the accessory parameter to a
  semi-classical conformal description and discuss the properties of
  the operators involved. For the radial equation, while the operators
  at the horizons have Liouville momenta proportional to the entropy
  intake, we find that spatial infinity is described by a Whittaker
  operator.
\end{abstract}

\keywords{Conformal Field Theory, Painlevé Transcendents, Black Holes}

\maketitle

\section{Introduction}

The Kerr black hole is described by the metric, in Boyer-Lindquist
coordinates \cite{Wald:1984}:
\begin{equation}
  ds^{2} = -\left(1-\frac{2Mr}{\Sigma}\right) dt^{2}-
  \frac{4Mar \sin^{2}\theta}{\Sigma}\,dt\,d\phi+
  \frac{\Sigma}{\Delta}dr^2+\Sigma d\theta^2 
  +\sin^{2}\theta
   \left(r^{2}+a^{2}+\frac{2Ma^{2}r\sin^{2}\theta}{\Sigma}\right)
  d\phi^{2}, 
\end{equation}
where 
\begin{equation}
  \Delta = r^{2}-2Mr+a^{2} = (r-r_{+})(r-r_{-}),\quad\quad
  \Sigma = r^{2}+a^{2}\cos^{2}\theta, \quad\quad
  a= \frac{J}{M},
  \label{eq:kerrparameters}
\end{equation}
referring to a solution of the four-dimensional, vacuum Einstein
equations which is asymptotically flat and has mass $M$ and angular
momentum $J=aM$. It has two event horizons at $r_\pm$ and the region
$r<r_+$ cannot affect causally the region $r>r_+$. The importance of
this solution to the development of general relativity and all
theories that generalize it can hardly be overestimated, since it is
shown to be the most general vacuum metric with mass and angular
momenta. Subsequent studies trying to reconcile its apparent
simplicity with the multitude of processes which can in principle
surround a black hole led to the concept of black hole entropy. The
microscopic description of the latter for generic black holes remains
an outstanding problem in theoretical physics.

However, the significance of the Kerr metric goes beyond formal
developments due to the various astrophysical applications of
phenomena in the Kerr background, especially the experimental
detection of the black hole ringdown after a black hole merging event
\cite{Abbott:2016blz}, 
as well as the recent image of a supermassive black hole whose shadow
region \cite{Akiyama:2019cqa} gives strong evidence of the existence
of an event horizon. As 
a matter of fact, both experiments are interpreted as a direct
evidence of a Kerr black hole, and from the raw data the black hole
parameters, such as $M$ and $a$, are measured.

All of these phenomena underscore the importance of the study of
fluctuations of the Kerr metric. Their evolution is described by the
linearized Einstein 
equations, with the metric fluctuations decomposed into a linear (spin
$0$), vector (spin $1$) and tensor (spin $2$) parts. The resulting 
partial differential equations are linear, separable and the spin $s$ solution can
be written as a sum of solutions of two ordinary differential
equations,
\begin{gather}
  \frac{1}{\sin\theta}\frac{d}{d\theta}\left[\sin \theta
    \frac{dS}{d\theta}\right]+
  \left[a^{2}\omega^{2}\cos^{2}\theta-2a\omega s \cos\theta -
    \frac{(m+s\cos\theta)^{2}}{\sin^{2}\theta}+s+\lambda
  \right]S(\theta)=0,
  \label{eq:angulareq} \\
  \Delta^{-s}\frac{d}{dr}\left(\Delta^{s+1}\frac{dR(r)}{dr}\right)+
  \left(\frac{K^{2}(r)-2is(r-M)K(r)}{\Delta}+4is\omega
   r-{_s\lambda_{\ell,m}}-a^2\omega^2+2am\omega
  \right)R(r)=0,
  \label{eq:radialeq}
\end{gather}
where
\begin{equation}
  K(r)=(r^2+a^2)\omega-am,\quad\quad \Delta =
  r^2-2Mr+a^2=(r-r_+)(r-r_-),
\end{equation}
which are called the (vacuum) \textit{Teukolsky master equations}
\cite{Teukolsky:1973ha}. The
spin $1$ version can be seen to describe the coupling of the
electromagnetic field to the black hole, and one can make $s=1/2$ to
describe massless spinorial particles. The derivation and the behavior
of the solutions for the Teukolsky master equation are the subject of
many articles and monographs, \textit{e. g.} \cite{Chandrasekhar1983}. 

The equations \eqref{eq:angulareq} and \eqref{eq:radialeq} have been
studied for decades now, the problem of scattering and the quasinormal
modes being the main topics of interest. Quasinormal modes were dealt
with extensively in \cite{Berti:2005ys,Berti:2009kk}, and fast
numerical techniques exist to compute the spectra of
perturbations. From the analytical side, less is known about the
behavior of the solutions and, while asymptotic formulas for the angular
eigenvalue exist \cite{Berti:2005gp,Casals:2009zh}, not much can be said
about the analytic behavior of the spectrum of eigenmodes. A better analytic
grasp on these would be invaluable to the study of phenomena such as
stability, superradiance and degeneracy properties of the spectrum.

The purpose of this article is to study eigenmodes of the Teukolsky
master equation analytically, using the isomonodromy method and its
relation to classical conformal blocks. The application of the
isomonodromy method to black holes was developed from early extensions
of the WKB method using monodromy techniques
\cite{Motl:2003cd,Neitzke2007}, developed also in
\cite{Castro2013,Castro2013b}. In \cite{daCunha:2015ana}, the
isomonodromy symmetry was introduced, and the relation between the
latter and $c=1$ conformal blocks was pointed to give a formal
solution to the scattering problem. Parallel developments allowed for
eigenmodes expansions for scalar perturbations (of generic mass) to
the five-dimensional Kerr-AdS black hole \cite{Barragan-Amado:2018pxh}
and generic massless perturbations of the four-dimensional Kerr-dS
black hole \cite{Novaes:2018fry}. These ordinary differential
equations involved in these problems are Fuchsian, and the relation
between the connection and monodromy property of their solutions was
outlined in \cite{Novaes:2014lha}.

Generically, the relation between the parameters of a Fuchsian
equation and the monodromy properties of their solutions is the oldest
form of the Riemann-Hilbert problem. This consist of determine a
particular complex function from its singular behavior. Since the
inception of this problem, the solution of the Riemann-Hilbert problem
has been related to solutions of the classical Liouville
equation. Surprisingly, the \textit{quantum} version of Liouville which
provided the window to the procedural construction of these
solutions. In 2009, Alday, Gaiotto and Tachikawa \cite{Alday:2009aq}
conjectured that the correlation functions of conformal primaries in
quantum Liouville theory would be the same as the instanton partition
function of some four-dimensional supersymmetric Yang-Mills theories,
which were given by Nekrasov functions \cite{Nekrasov:2002qd}. The
relation was proven in \cite{Alba:2010qc}, through combinatorial
means solving recursion conditions on the representations of the
Virasoro algebra.

Liouville field theory is also related to the theory
of flat holomorphic connections, and the monodromy data of the latter
is encoded in the isomonodromic $\tau$ functions, of which the simplest
non-trivial examples where introduced by Jimbo, Miwa and Ueno in
\cite{Jimbo:1981aa,Jimbo:1981ab,Jimbo:1981ac}. These $\tau$ functions
have the Painlevé property \cite{Miwa:1981aa}, and the simplest
examples are guises of the six Painlevé transcendents, solutions of
ordinary differential non-linear equations of second order with
rational coefficients whose essential singularities are determined
from the equation itself, and the remaining singularities are single
poles. Following the Liouville field theory realization, expansions of
the sixth, fifth and third Painlevé transcendents were given in terms
of $c=1$ conformal blocks in \cite{Gamayun:2013auu}. The relation was
further explored in \cite{Iorgov:2014vla} and Fredholm determinant
expression for a generic class of Painlevé transcendents were given in
\cite{Gavrylenko:2016zlf}. The Fredholm determinant formulation allows
for faster numerical calculations, as well as a more direct contact
between usual applications of the Riemann-Hilbert problem, such as
those in matrix models, and conformal field theory methods. 

The overall program of phrasing perturbations of gravitational
backgrounds in terms of conformal blocks has a holographic flavor
which was also explored by a number of authors, see, for instance,
\cite{Hijano:2015zsa}. The purpose, however, can be understood to be
what conditions in the purported dual theory one gets from the
integrable structure of the gravitational perturbations, rather than
the other way around. The latter program of framing the dual theory
from the particular integrable structure of the perturbations is bound
to be valuable for asymptotically flat spaces, where the dual theory
is not so clear cut. In the five-dimensional Kerr-AdS case studied in
\cite{Amado:2017kao}, such conditions were observed to arise from a 
\textit{unitary} conformal field theory. The relation between the 
$c=1$ blocks used to construct the relation between monodromy
parameters and scattering coefficients and the semiclassical
prescription outlined in \cite{Nekrasov:2011bc,Litvinov:2013sxa}
remains mysterious and may yet shed light in a true quantum
description of the black hole states. 

In this article we will carry on the analysis of the quasinormal modes
of the Teukolsky master equation by exploring the interpretation of
the differential equations involved with conformal blocks. We will see
that the relevant conformal blocks are irregular, as studied by
\cite{Gaiotto:2009ma} and \cite{Nagoya:2010yb}, and the relevant
Painlevé transcendent the fifth type \cite{Nagoya:2015cja}. We will
see that, while the $c=1$ blocks give an analytic solution to the
accessory parameter problem, exponentiation also allows for a
description in terms of semi-classical irregular conformal blocks.

The paper is structured as follows. In Sec. \ref{sec:preamble} we will
introduce the monodromy data associated to the relevant differential
equation, as well as phrase the connection problem in terms of the
isomonodromic $\tau$ function. In Sec. \ref{sec:angular} we will apply
the method to the angular differential equation \eqref{eq:angulareq} and
obtain expansions for the spin-weighted spheroidal harmonic
eigenvalue. In Sec. \ref{sec:radial} we will revise the conformal
block version of the construction, and apply it to the radial equation
\eqref{eq:radialeq}, interpreting the semiclassical conformal block as a
correlation function of an unitary theory and using the method to
obtain some quasinormal modes. We close by remarking on the future
prospects in Sec. \ref{sec:discussion}.

\section{Preamble: the confluent Heun equation}
\label{sec:preamble}

Both equations \eqref{eq:angulareq} and \eqref{eq:radialeq} can be
brought to the confluent Heun canonical form:
\begin{equation}
  \frac{d^{2}y}{dz^{2}}+\left[\frac{1-\theta_{0}}{z}+\frac{1-\theta_{t_{0}}}{z-t_{0}}
    \right] \frac{dy}{dz}+
    \left[-\frac{1}{4}+\frac{\theta_\infty}{2z}-\frac{t_0c_{t_{0}}}{z(z-t_{0})}
    \right]y(z)=0,
    \label{eq:confluentheun}
\end{equation}
The differential equation \eqref{eq:confluentheun} has $3$ singular
points: two regular at $z=0$ and $z=t_0$ and an irregular singular point of
Poincaré rank $1$ at
$z=\infty$. Series expansions for the solutions $y(z)$ at the regular points
can be obtained from the Frobenius method. The point at infinity is
trickier, because the solutions present the \textit{Stokes
  phenomenon}: convergence is conditional to sectors of the complex
plane, depending on the direction one takes the limit $z\rightarrow
\infty$. 

Near a regular singular point $z_i$, the Frobenius method allows us,
in general, to construct two solutions, whose local behavior is
\begin{equation}
  y_i^{\pm}(z)=(z-z_i)^{\frac{1}{2}\alpha_i\pm\frac{1}{2}\theta_i}
  (1+{\cal O}(z-z_i))
\end{equation}
which we will call the local Frobenius solutions at $z=z_i$. In
general, one given Frobenius solution at $z=z_i$ will be expressed as
a linear combination of the Frobenius solutions constructed at a
different point $z=z_j$. For a particular set of parameters in the
differential equation \eqref{eq:confluentheun}, namely discrete values
of the accessory parameter $c_{t_0}$, there will be a solution which
has definite behavior at \textit{both} $z=z_i$ and $z=z_j$, for instance:
\begin{equation}
  y(z) = \begin{cases}
    (z-z_i)^{\frac{1}{2}\alpha_i+\frac{1}{2}\theta_i}(1+{\cal
      O}(z-z_i)), & z\rightarrow z_i; \\
     (z-z_j)^{\frac{1}{2}\alpha_j+\frac{1}{2}\theta_j}(1+{\cal
       O}(z-z_j)), & z\rightarrow z_j.
    \label{eq:eigenproblem}
 \end{cases}
\end{equation}
Finding the (discrete) values of $c_{t_0}$ for which such a $y(z)$ exists will be
referred to as the \textit{eigenvalue problem}. The formulation of the
eigenvalue problem at the irregular singular point $z=\infty$ is a bit
more complicated and will be dealt with later.

Let us briefly describe the solution to the eigenvalue problem
proposed in \cite{daCunha:2015ana}. The second order differential
equation \eqref{eq:confluentheun} can be cast as a first order
matrix equation:
\begin{equation}
  \frac{d\Phi}{dz}\Phi^{-1}(z)=A(z)=
  \frac{1}{2}\begin{pmatrix} 1 & 0 \\ 0 & -1 \end{pmatrix}
  +\frac{A_0}{z}+\frac{A_{t}}{z-t}
  =\frac{1}{2}\sigma_3+\frac{A_0}{z}+\frac{A_t}{z-t},
  \label{eq:matrixsystem}
\end{equation}
where we introduced the \textit{fundamental matrix} of solutions
$\Phi(z)$:
\begin{equation}
  \Phi(z)=\begin{pmatrix}
    y_1(z) & y_2(z) \\
    w_1(z) & w_2(z)
  \end{pmatrix},
  \label{eq:fundamentalmatrix}
\end{equation}
with $y_{1,2}(w)$ satisfying our original equation
\eqref{eq:confluentheun} and $w_{1,2}(z)$ related to $y_{1,2}(z)$ by
differentiation and multiplication by a rational function:
\begin{equation}
  w_i(z) =
  \frac{1}{A_{12}(z)}\left(\frac{dy_i}{dz}-A_{11}(z)y_i(z)\right).
\end{equation}
We note that any two solutions of \eqref{eq:matrixsystem} are related
by right multiplication. We also note that one can change the value of
$\alpha_i$ at will by multiplication of the solution
\eqref{eq:fundamentalmatrix} by a factor
$\prod_i(z-z_i)^{\frac{1}{2}\alpha_i}$, with exception of the singular
point at infinity. 

The basis of the method is to see the parameter $t$ in
\eqref{eq:matrixsystem} as a gauge parameter in the space of flat
holomorphic connections $A(z,t)$, and to recover the differential
equation \eqref{eq:confluentheun} as we take $t$ to $t_0$. The
usefulness of this deformation stems from the fact that we can
translate conditions such as the quantization condition
\eqref{eq:boundary} in terms of gauge-invariant properties of
\eqref{eq:matrixsystem}, called \textit{monodromy data}. 

\subsection{Monodromy data}

Let us first describe the latter. The monodromy data associated to the
matrix of solution $\Phi(z)$ of \eqref{eq:matrixsystem} is its
behavior under analytical continuation around the singular points:
\begin{equation}
  \Phi((z-z_0)e^{2\pi i}+z_0)=\Phi(z)M_{z_0},
  \label{eq:monodromymatrix}
\end{equation}
which defines the monodromy matrix as the decomposition of the
analytic continuation of each of the solutions in terms of
themselves. As defined above, the matrices $M_i$ are independent of
the homotopy class of the curve we choose for analytic
continuation. The matrices $M_i$ are also independent on the sum of
the indicial exponents at each singular point, the $\alpha_i$ in
\eqref{eq:eigenproblem}, due to the fact that these can be changed by
multiplication of a scalar function.

For the irregular singular point $z=\infty$ there
is a subtlety, due to the Stokes phenomenon. Let us follow
\cite{Jimbo:1982aa} (see also \cite{Andreev:2000aa} -- and define
sectors ${\mathcal S}_k$ as 
\begin{equation}
  {\mathcal S}_k=\left\{ z\in \mathbb{C}\ , \  -\frac{1}{2}\pi
    +(k-2)\pi< \arg z < \frac{3}{2}\pi+(k-2)\pi\right\},\quad\quad
  k\in\mathbb{Z}
\end{equation}
in each the asymptotic solution for \eqref{eq:matrixsystem} is
\begin{equation}
  \Phi_k(z)=\left(\mathbbold{1}+{\mathcal O}(z^{-1})\right)
  \exp\left[\frac{1}{2}\sigma_3z+\frac{1}{2}((\hat{\theta}_0+
    \hat{\theta}_t) 
    \mathbbold{1}-\hat{\theta}_\infty\sigma_3)\log z\right],
  \quad\quad  z\rightarrow \infty,\ z\in {\mathcal S}_k,
  \label{eq:asympphik}
\end{equation}
where $\theta_i$ are defined as
\begin{equation}
  \hat{\theta}_0=\Tr A_0,\quad\quad
  \hat{\theta}_t=\Tr A_t,\quad\quad
  \hat{\theta}_\infty=-\Tr\left[\sigma_3(A_0+A_t)\right].
\end{equation}
The analytic continuation of the solution $\Phi_k(z)$ can be now
described as the connection between $\Phi_k(z)$ in different sectors:
\begin{equation}
  \Phi_{k+1}(z)=\Phi_k(z)S_k,
\end{equation}
where $S_k$ are the \textit{Stokes matrices}. By \eqref{eq:asympphik},
\begin{equation}
  S_{k+2}=e^{i\pi\hat{\theta}_\infty\sigma_3}
  S_ke^{-i\pi\hat{\theta}_\infty\sigma_3},
\end{equation}
so only two of the Stokes matrices are independent. It can be checked
from the discussion that they have the structure
\begin{equation}
  S_{2k}=\begin{pmatrix} 1 & s_{2k} \\ 0 & 1 \end{pmatrix},
  \quad\quad
  S_{2k+1}=\begin{pmatrix} 1 & 0 \\  s_{2k+1} & 1 \end{pmatrix},
  \quad\quad k\in\mathbb{Z},
\end{equation}
where the parameters $s_{2k},s_{2k+1}$ are called \textit{Stokes
  multipliers}. It is customary to define the monodromy matrix at
$z=\infty$ in the sector $k=2$:
\begin{equation}
  \Phi_2(z e^{-2\pi i})=\Phi_2(z) M_\infty(k=2)=\Phi_2(z)M_\infty,
\end{equation}
with the corresponding matrices for generic $k$ defined through the
recursion $M_\infty(k+1)=S_k^{-1}M_\infty(k)S_k$. The monodromy matrix
$M_\infty$ can be obtained from the Stokes matrices by
\begin{equation}
  M_\infty = S_2 e^{i\pi\hat{\theta}_\infty\sigma_3}S_1
  \label{eq:monoinfty}
\end{equation}
and satisfies the relation
\begin{equation}
  M_\infty M_t M_0=\mathbbold{1}.
  \label{eq:cyclic}
\end{equation}

With these definitions, we define the monodromy data $\rho$ associated
to the matrix equation \eqref{eq:matrixsystem} as the basis
independent data in the matrices $M_i$:
\begin{equation}
  \rho = \{ \hat{\theta}_0,\hat{\theta}_t,\hat{\theta}_\infty;s_1,s_2\}.
\end{equation}
It will be convenient to define the trace of $M_\infty$ as an
independent parameter
\begin{equation}
  2\cos\pi\hat{\sigma} = \Tr M_\infty = 2\cos\pi\hat{\theta}_\infty
  +s_1s_2e^{-i\pi\hat{\theta}_\infty}.
  \label{eq:sigma}
\end{equation}

\subsection{Connection matrix and the quantization condition}

We can now phrase the eigenvalue problem \eqref{eq:eigenproblem}
in terms of monodromy data. Let us choose the fundamental solution at
$z=0$, $\Phi(z;z_0=0)$ with $y_1(z)$ and $y_2(z)$ in
\eqref{eq:fundamentalmatrix} constructed using the Frobenius method at
$z=0$.
\begin{equation}
  \Phi(z;0)=\left(\mathbbold{1}+{\mathcal O}(z)\right)
  \exp\left[\left(\frac{1}{2}\alpha_0\mathbbold{1}+\frac{1}{2}
    \hat{\theta}_0\sigma_3\right)\log z
  \right],
\end{equation}
where $\tfrac{1}{2}(\alpha_0\pm\hat{\theta}_0)$ are the eigenvalues of
$A_0$.  It is clear that the monodromy matrix around $z_0=0$ for this
basis is diagonal:
\begin{equation}
  \Phi(z e^{2\pi i};0)=\Phi(z;0) e^{i\pi\alpha_0}e^{i\pi \hat{\theta}_0\sigma_3}.
\end{equation}
The $\alpha_0$, abelian part of the monodromy can be removed by a
``s-homotopic transformation'' like \eqref{eq:shomotopic} and can be
taken to be zero. We therefore have that, in this basis of solutions
$M_0=e^{i\pi\hat{\theta}_0\sigma_3}$. The monodromy around $z=t$ is
likewise diagonal with the fundamental solution $\Phi(z;z_0=t)$, but
in terms of $\Phi(z,z_0=0)$ above
\begin{equation}
  \Phi((z-t)e^{2\pi i}+t;0)=\Phi(z;0)C^{-1}_{t0}e^{\pi
    i\hat{\theta}_t\sigma_3}C_{t0},
  \quad\quad\text{where}\quad C_{t0}=\Phi(z;t)^{-1}\Phi(z;0),
\end{equation}
is called the \textit{connection matrix} between the singular points
at $z=0$ and $z=t$. 

Now, we can see that if the parameters in the matrix system
\eqref{eq:matrixsystem} are such that the conditions
\eqref{eq:boundary} are satisfied, then the connection matrix
$C_{t0}$ is either lower triangular or upper triangular. Simple
algebra shows that, if this is the case, then
\begin{equation}
  \Tr M_0M_t=2\cos\pi(\hat{\theta}_0+\hat{\theta}_t).
\end{equation}
It can be checked that the converse is also true: if this trace
property is satisfied, then $C_{t0}$ is either lower or upper
triangular. This is a condition to be satisfied when $\lambda$
corresponds to the angular eigenvalue. Using the property
\eqref{eq:cyclic}, we have $\Tr M_0M_t=\Tr M_\infty^{-1}$, and, by the
definition of $\sigma$ above 
\eqref{eq:sigma}, we arrive at
\begin{equation}
  \cos\pi(\hat{\theta}_0+\hat{\theta}_t)=\cos\pi\hat{\sigma},
  \Longrightarrow
  \hat{\sigma}(\lambda_\ell)=\hat{\theta}_0+\hat{\theta}_t+2j,
  \quad\quad
  j \in \mathbb{Z}, 
  \label{eq:quantization}
\end{equation}
where we underscored the dependence of the $\hat{\sigma}$ parameter on
$\lambda$, but in fact it depends on all parameters in
\eqref{eq:confluentheun}.

The condition \eqref{eq:quantization} does not provide a full solution
of the system, however, because it may involve non-normalizable
solutions of the differential equation \eqref{eq:confluentheun}. In
our applications below, it will be clear from the context which values
of $\theta_i$ lead to the proper modes. 

\subsection{The $\tau$ function and Painlevé V system}

To calculate $\sigma$ as a function of the differential equation
parameters is a version of the Riemann-Hilbert problem, whose solution
we will make use of. The idea goes back to the theory of isomonodromic
deformations as introduced by
\cite{Jimbo:1981aa,Jimbo:1981ab,Jimbo:1981ac}, and is based on
interpreting $t$ as a gauge parameter. If we accompany
\eqref{eq:matrixsystem} by its Lax pair:
\begin{equation}
  \frac{\partial\Phi}{\partial t}[\Phi(z,t)]^{-1} = -\frac{A_t}{z-t}
\end{equation}
the existence of the mixed derivative
$\partial_z\partial_t\Phi=\partial_t\partial_z\Phi$ requires that
$A_0$ and $A_t$ satisfy the \textit{Schlesinger equations}:
\begin{equation}
  \frac{\partial A_0}{\partial t}=\frac{1}{t}[A_t,A_0],\quad\quad
  \frac{\partial A_t}{\partial t}=-\frac{1}{t}[A_t,A_0]-\frac{1}{2}
  [A_t,\sigma_3],
  \label{eq:schlesinger}
\end{equation}
whose solution gives a one-parameter family of matrix systems with
different values of $t$ but the same monodromy data. Since $A_0$ and
$A_t$ are now arbitrary, let us consider the generic differential
equation satisfied by the first row of $\Phi(z)$ in
\eqref{eq:matrixsystem}
\begin{gather}
  \frac{d^2y}{dz^2}+p(z)\frac{dy}{dz}+q(z)y=0, \\[5pt]
  p(z)=\frac{1-\hat{\theta}_0}{z}+
  \frac{1-\hat{\theta}_t}{z-t}-\frac{1}{z-\lambda}, 
  \quad\quad 
  q(z)=-\frac{1}{4}+\frac{\hat{\theta}_\infty-1}{2z}-
  \frac{tc_t}{z(z-t)}+\frac{\lambda\mu}{z(z-\lambda)},
\label{eq:deformedode}
\end{gather}
where $\lambda$ is the root of $A_{12}(z)$ and
$\mu=A_{11}(z=\lambda)$. $c_t$ is related to $\lambda$
and $\mu$ by 
\begin{equation}
  \mu^2-\left[\frac{\hat{\theta}_0}{\lambda}+
    \frac{\hat{\theta}_t-1}{\lambda 
      -t}\right]\mu+\frac{\hat{\theta}_\infty-1}{2\lambda}-
  \frac{tc_t}{\lambda(\lambda-t)}= 
  \frac{1}{4}.
  \label{eq:algebraicconstraint}
\end{equation}
The algebraic condition \eqref{eq:algebraicconstraint} tells us that
the singularity at $z=\lambda$ in \eqref{eq:deformedode} is an
apparent one: the indicial equation gives integer exponents $0$ and
$2$, and there is no logarithmic behavior due to
\eqref{eq:algebraicconstraint}. The monodromy matrix around
$z=\lambda$ is then trivial. The Schlesinger equations induce a
flow to $\lambda$ and $\mu$, and the corresponding differential
equation for $\lambda$ is equivalent to the Painlevé V transcendent.

The family of isomonodromic connections will include our original
equation \eqref{eq:confluentheun} if
\begin{equation}
  \hat{\theta}_0=\theta_0,\quad\quad
  \hat{\theta}_t=\theta_{t_0}-1,\quad\quad
  \hat{\theta}_\infty=\theta_\infty+1,\quad\quad
  \lambda(t_0)=t_0,\quad\quad
  \mu(t_0)=-\frac{c_{t_0}}{\theta_{t_0}-1},
  \label{eq:initialgarnier}
\end{equation}
and note that, per \eqref{eq:sigma}, $\hat{\sigma}=\sigma-1$. 
These conditions are more conveniently written in terms of the
Jimbo-Miwa-Ueno (JMU) $\tau$ function
\begin{equation}
  \frac{d}{d t}\log\tau(\rho;t)=\frac{1}{2}\Tr
  \sigma_3A_t+\frac{1}{t} 
  \Tr(A_0-\tfrac{1}{2}\hat{\theta}_0\mathbbold{1})
  (A_t-\tfrac{1}{2}\hat{\theta}_t\mathbbold{1}),
  \label{eq:taufunction}
\end{equation}
where we left explicitly the dependence of the JMU $\tau$ function on
the monodromy data $\rho$ due to its expansions \cite{Jimbo:1982aa},
\cite{Gamayun:2013auu}. Therefore,
\eqref{eq:initialgarnier} is
\begin{equation}
  \frac{d}{d t}\log\tau(\hat{\rho};t_0)= c_{t_0}+\frac{\hat{\theta}_0
    \hat{\theta}_t}{2t_0},\quad\quad
  \frac{d}{dt}t\frac{d}{dt}\log\tau(\hat{\rho};t_0)+\frac{\hat{\theta}_t}{2}=0.
  \label{eq:tauconditions}
\end{equation}
The second condition \eqref{eq:tauconditions} stems from the second
derivative of the $\tau$ function, calculated using the Schlesinger
equations and imposing \eqref{eq:initialgarnier}. The left hand side
can be related through the \textit{Toda equation}
\cite{Okamoto:1987aa} to a product of $\tau$ functions:
\begin{equation}
  \frac{d}{dt}t\frac{d}{dt}\log\tau(\hat{\rho};t_0)+\frac{\hat{\theta}_t}{2}
  = K_V\frac{\tau(\hat{\rho}^+;t)\tau(\hat{\rho}^-;t)}{\tau^2(\hat{\rho};\tau)},
  \label{eq:todaequation}
\end{equation}
where $K_V$ is independent of $t$ and the $\rho^\pm$ are related to
$\rho$ by simple shifts: 
\begin{equation}
  \hat{\rho}^\pm=\{\hat{\theta}_0,\hat{\theta}_t\pm 1,
  \sigma\pm 1, \hat{\theta}_\infty\mp 1;s_1,s_2\}.
  \label{eq:schlesingershift}
\end{equation}
Miwa's theorem \cite{Miwa:1981aa} tells us that $\tau$ defined by
\eqref{eq:taufunction} is analytic in $t$ except at the critical
points $t=0$ and $t=\infty$. Therefore either factor of the numerator
in \eqref{eq:todaequation} has to vanish.

The proof of \eqref{eq:todaequation} is straightforward, from a
fundamental solution $\Phi(z)$ one defines the derived solutions
\begin{equation}
  \Phi^\pm(z) = \exp[p^\pm \sigma^\mp] \begin{pmatrix}
    (z-t)^{\pm 1} & 0 \\ 0 & 1 \end{pmatrix}
  \exp[q^\pm\sigma^\pm]\Phi(z),
\end{equation}
where $\sigma^+=\begin{pmatrix} 0 & 1 \\ 0 & 0 \end{pmatrix}$ and
$\sigma^-=\begin{pmatrix} 0 & 0 \\ 1 & 0 \end{pmatrix}$  are nilpotent
combinations of Pauli matrices. Given 
$\Phi^\pm(z)$, one can establish the Toda equation \eqref{eq:todaequation}
by comparing the corresponding expressions for each $\tau$ function
\eqref{eq:taufunction}, and choosing $p^\pm$ and $q^\pm$ in order to
keep the form of the new connection, defined through
\eqref{eq:matrixsystem}, maintain the partial fraction form at $z=t$
and $z=\infty$. It is clear that the monodromy data of  $\Phi^\pm(z)$
are related to that of $\Phi(z)$ by
\eqref{eq:schlesingershift}. Further algebraic manipulation 
shows that 
\begin{gather}
  \frac{d}{dt}\log\frac{\tau(\rho^+;t)}{\tau(\rho;t)} =
  -\frac{1}{2}-\frac{\lambda}{t}\left(\mu-\frac{1}{2}\right)
  +\frac{\lambda}{t(\lambda-t)}\hat{\theta}_t \\
  \frac{d}{dt}\log\frac{\tau(\rho^-;t)}{\tau(\rho;t)} =
  \frac{1}{2}-\frac{(\lambda-t)\left(\mu-\frac{1}{2}\right)-
    \frac{1}{2}(\hat{\theta}_0+\hat{\theta}_t-\hat{\theta}_\infty)}{
    \lambda\left(\mu-\frac{1}{2}\right)-\frac{1}{2}(\hat{\theta}_0
    +\hat{\theta}_t-\hat{\theta}_\infty)}\left(
    \frac{\lambda}{t}\left(\mu-\frac{1}{2}\right)-\frac{\hat{\theta}_0}{t}
  \right).
\end{gather}
Given that the first line has a divergent limit $\lambda\rightarrow
t$, we conclude that we can substitute the second condition in
\eqref{eq:tauconditions} by the simpler one
\begin{equation}
  \tau(\rho;t_0)=0,
  \label{eq:zerotau}
\end{equation}
where the monodromy data is that of \eqref{eq:confluentheun}:
\begin{equation}
  \rho=\{\theta_0,\theta_{t_0},\theta_\infty;s_1,s_2\},
\end{equation}
whereas, in terms of $\rho$, the first condition in
\eqref{eq:tauconditions} is given by
\begin{equation}
  c_{t_0}=\frac{d}{dt}\log\tau(\rho^-;t_0)-\frac{\theta_0(\theta_{t_0}-1)}{2t_0}.
  \label{eq:difflogtau}
\end{equation}
with the shift in $\rho^-$ analogous to that of $\hat{\rho}$ above.

\subsection{The Nekrasov expansion}
\label{sec:nekrasov}

In this section we are going to drop the ``hatted''  notation in
order not to overburden the formulas.
The Nekrasov expansion of the Painlevé V $\tau$-function is given by
\cite{Gamayun:2013auu}
\begin{equation}
  \tau(\rho;t) =
  \sum_{n\in\mathbb{Z}}C_{V}(\vec{\theta},
  \sigma+2n)s_V^{n}t^{\tfrac{1}{4}(\sigma+2n)^{2}-\tfrac{1}{4}(\theta_0^2+\theta_t^2)}
  \mathcal{B}_{V}(\vec{\theta};\sigma+2n;t).
  \label{eq:nekrasov}
\end{equation}
Here $\rho=\{\theta_0,\theta_t,\theta_\infty;\sigma,s_V\}$ is the
monodromy data. The definition of the parameter $\sigma$ in terms of
the Stokes parameters is given by \eqref{eq:sigma}, and we will discuss the
parameter $s_V$ below. The function $\mathcal{B}_V$ is analytic near
$t=0$ and closely related to the irregular conformal blocks of the
first kind \cite{Gaiotto:2009ma,Nagoya:2015cja}. It is based on the
Nekrasov expansion, which a scalar function associated to a pair of
Young diagrams $\lambda,\mu$, a complex parameter $b$, as well as a
complex number $\alpha$: 
\begin{equation}
  Z_{\lambda,\mu}(\alpha)=\prod_{(i,j)\in \lambda}
  (\alpha+b^{-1}a_\lambda(i,j)-b(l_\mu(i,j)+1))
  \prod_{(i',j')\in \mu}
  (\alpha-b^{-1}(a_\mu(i',j')+1)+bl_\lambda(i',j')),
\end{equation}
where $a_\lambda(i,j)$ and $l_\lambda(i,j)$ are respectively the
arm-length and the leg-length of the box $(i,j)$ in the diagram
$\lambda$. The parameter $b$ is related to the central charge of the
Virasoro algebra by $c=1+6Q^2=1+6(b+b^{-1})^2$
\eqref{eq:liouvilledefs}. As it can be checked in
\cite{Lisovyy:2018mnj}, the expansion of irregular conformal block of
the first kind is given by  
\begin{equation}
  \mathcal{B}\left(P_\infty;P_\sigma;
    \substack{P_{t} \\ P_{0}};t\right)=
  t^{\Delta_\sigma-\Delta_0-\Delta_t}e^{-i(\tfrac{Q}{2}-iP_t)t}
  \sum_{\lambda,\mu\in
    \mathbb{Y}} \mathcal{B}_{\lambda,\mu}(\vec{P};P_\sigma)
  t^{|\lambda|+|\mu|}
  \label{eq:confluentconformal}   
\end{equation}
where $\Delta_i = \tfrac{Q^2}{4}+P_i^2$
and $\mathcal{B}_{\lambda,\mu}$ is given by ratios of Nekrasov
functions
\begin{multline}
  \mathcal{B}_{\lambda,\mu}(\vec{P};P_\sigma)=
  \frac{Z_{\lambda,\emptyset}(\tfrac{Q}{2}-i(P_\infty-P_\sigma))
    Z_{\mu,\emptyset}(\tfrac{Q}{2}-i(P_\infty+P_\sigma))}{
    Z_{\lambda,\lambda}(0)Z_{\mu,\mu}(0)Z_{\lambda,\mu}(2iP_\sigma)
    Z_{\mu,\lambda}(-2iP_\sigma)}\\
  \times \prod_{\epsilon=\pm}
    Z_{\lambda,\emptyset}(\tfrac{Q}{2}+i(P_t+\epsilon
    P_0+P_\sigma))
    Z_{\emptyset,\mu}(\tfrac{Q}{2}+i(P_t+\epsilon
    P_0-P_\sigma)).
    \label{eq:genericnekrasov}
\end{multline}

As stated in \cite{Gamayun:2013auu}, the expansion of the
$\tau$-function for the Painlevé V near $t=0$ is given in terms of $c=1$
irregular conformal blocks. These are obtained taking $b=\sqrt{-1}$ --
and therefore $Q=0$ -- in the expressions above, as well as setting
the parameters $P_i$ to the monodromy parameters:
\begin{equation}
  P_0=\frac{\theta_0}{2},\quad\quad P_t=\frac{\theta_t}{2},
  \quad\quad P_\infty = \frac{\theta_\infty}{2},
  \quad\quad P_\sigma = \frac{\sigma}{2}.
\end{equation}
Coming back to \eqref{eq:nekrasov}, one can recognize in the ${\cal
  B}_V$ expansion the terms of the same functions
$\mathcal{B}_{\lambda,\mu}$ appearing in the expansion of the
irregular conformal blocks \eqref{eq:genericnekrasov}:
\begin{equation}
  \mathcal{B}_{V}(\vec{\theta},\sigma;t)=
  e^{-\frac{\theta_{t}}{2}t}\sum_{\lambda, \mu \in \mathbb{Y}}
  \mathcal{B}_{\lambda,\mu}(\vec{\theta},\sigma)t^{|\lambda|+ |\mu|}, 
  \label{eq:irrcb}
\end{equation}
where, again, the sum runs over all pairs of Young diagrams $(\lambda,\mu)$,
with each coefficient in the series given by the appropriate reduction
of \eqref{eq:genericnekrasov}:
\begin{multline}
  \mathcal{B}_{\lambda, \mu}(\vec{\theta},\sigma) =\prod_{\substack{
      \lambda \in \mathbb{Y}}}\frac{(2(i-j)+\sigma-\theta_\infty)((
    \sigma+\theta_t+2(i-j))^2-\theta_0^2)}{8h_\lambda^2(i,j)
    (l_\lambda(i,j)+a_\mu(i,j)+1+\sigma)^2} \\
  \times\prod_{\substack{
      \mu \in \mathbb{Y}}}
  \frac{(2(i-j)-\sigma-\theta_\infty)
    ((\theta_t-\sigma+2(i-j))^2-\theta_0^2)}{8 h_\mu^2(i,j)
    (a_\lambda(i,j)+l_\mu(i,j)+1-\sigma)^2},
\end{multline}
and the the hook lenght is defined by
$h_{\lambda}(i,j)=a_\lambda(i,j)+l_\lambda(i,j)+1$. The structure
constants $C_{V}$ in \eqref{eq:nekrasov} are rational products of
Barnes functions  
\begin{equation}
  C_{V}(\vec{\theta}, \sigma) = {\mathcal N}(\vec{\theta},\sigma)
  {\mathcal N}(\vec{\theta},-\sigma)
\end{equation}
where
\begin{equation}
  {\mathcal N}(\vec\theta,\sigma)=
  \frac{G(1+\tfrac{1}{2}(\sigma-\theta_\infty))
    G(1+\tfrac{1}{2}(\theta_t+\theta_0+\sigma))
    G(1+\tfrac{1}{2}(\theta_t-\theta_0+\sigma))}{ G(1+\sigma)}.
\end{equation}
where the Barnes function $G(z)$ is defined by functional equation
$G(1+z)=\Gamma(z)G(z)$ plus some convexity requirements. The
functional equation is its only property required to recover the
results in this paper. 

\subsection{Monodromy matrices}
\label{sec:monomatrices}

The parameter $s_V$ in \eqref{eq:nekrasov} has a geometrical
interpretation in terms of the monodromy data. Following
\cite{Jimbo:1982aa,Andreev:2000aa}, we will introduce an explicit
representation for the monodromy matrices. Let
\begin{equation}
  M_0=C^{-1}_0e^{i\pi\theta_0\sigma_3}C_0,\quad\quad
  M_t=C^{-1}_te^{i\pi\theta_t\sigma_3}C_t,\quad\quad
  M_\infty = S_2 e^{i\pi\theta_\infty\sigma_3}S_1 .
\end{equation}
The connection matrices $C_0$ and $C_t$ allow the following
parametrization:
\begin{gather}
  D_t C_t D =
  \begin{pmatrix}
       \frac{\Gamma(1-\sigma)\Gamma(-\theta_t)}{
      \Gamma(-\frac{1}{2}(\theta_t+\theta_0+\sigma))
      \Gamma(-\frac{1}{2}(\theta_t-\theta_0+\sigma))} &
       \frac{\Gamma(1+\sigma)\Gamma(-\theta_t)}{
      \Gamma(-\frac{1}{2}(\theta_t+\theta_0-\sigma))
      \Gamma(-\frac{1}{2}(\theta_t-\theta_0-\sigma))} \\
      \frac{e^{-\pi i\theta_t}\Gamma(1-\sigma)\Gamma(\theta_t)}{
      \Gamma(1+\frac{1}{2}(\theta_t+\theta_0-\sigma))
      \Gamma(1+\frac{1}{2}(\theta_t-\theta_0-\sigma))} &
      \frac{e^{-\pi i \theta_t}\Gamma(1+\sigma)\Gamma(\theta_t)}{
      \Gamma(1+\frac{1}{2}(\theta_t+\theta_0+\sigma))
      \Gamma(1+\frac{1}{2}(\theta_t-\theta_0+\sigma))} 
  \end{pmatrix}
  \kappa^{-\tfrac{1}{2}\sigma_3}
  C_\infty, \label{eq:connectiont} \\
    D_0 C_0 D =
\begin{pmatrix}
    \frac{e^{\frac{1}{2}\pi i(\theta_t-\theta_0+\sigma)}
  \Gamma(1-\sigma)\Gamma(-\theta_0)}{
      \Gamma(1+\frac{1}{2}(\theta_t-\theta_0-\sigma))
      \Gamma(-\frac{1}{2}(\theta_t+\theta_0+\sigma))} &
    \frac{e^{\frac{1}{2}\pi i(\theta_t-\theta_0-\sigma)}
      \Gamma(1+\sigma)\Gamma(-\theta_0)}{
      \Gamma(1+\frac{1}{2}(\theta_t-\theta_0+\sigma))
      \Gamma(-\frac{1}{2}(\theta_t+\theta_0-\sigma))} \\
    \frac{e^{\frac{1}{2}\pi i(\theta_t+\theta_0+\sigma)}
      \Gamma(1-\sigma)\Gamma(\theta_0)}{
      \Gamma(1+\frac{1}{2}(\theta_t+\theta_0-\sigma))
      \Gamma(-\frac{1}{2}(\theta_t-\theta_0+\sigma))} &
    \frac{e^{\frac{1}{2}\pi i(\theta_t+\theta_0-\sigma)}
      \Gamma(1+\sigma)\Gamma(\theta_0)}{
      \Gamma(1+\frac{1}{2}(\theta_t+\theta_0+\sigma))
      \Gamma(-\frac{1}{2}(\theta_t-\theta_0-\sigma))}
  \end{pmatrix}
  \kappa^{-\tfrac{1}{2}\sigma_3}
  C_\infty,
\end{gather}
with $D_t$, $D_0$ and $D$ diagonal matrices and
\begin{equation}
  C_\infty = \begin{pmatrix}
    -e^{- i\frac{\pi}{2}(\sigma+\theta_\infty)}\frac{\Gamma(1-\sigma)}{
      \Gamma(-\frac{1}{2}(\sigma-\theta_\infty))} &
    -\frac{\Gamma(1-\sigma)}{\Gamma(1-\frac{1}{2}(\sigma+\theta_\infty))}
      \\
     e^{i\frac{\pi}{2}(\sigma-\theta_\infty)}\frac{\Gamma(1+\sigma)}{
      \Gamma(\frac{1}{2}(\sigma+\theta_\infty))} &
    \frac{\Gamma(1+\sigma)}{\Gamma(1+\frac{1}{2}(\sigma-\theta_\infty))}
  \end{pmatrix}.
  \label{eq:connectioninfty}
\end{equation}
The $\kappa$ parameter in the monodromy matrix is related to the $s_V$
parameter in the Nekrasov expansion \eqref{eq:nekrasov} by a string of
gamma functions
\begin{equation}
  \kappa =
  \frac{\Gamma^2(1-\tilde{\sigma})}{\Gamma^2(1+\tilde{\sigma})}
  \frac{
    \Gamma(1+\tfrac{1}{2}(\tilde{\sigma}-\theta_\infty))
    \Gamma(1+\tfrac{1}{2}(\theta_t+\theta_0+\tilde{\sigma}))
    \Gamma(1+\tfrac{1}{2}(\theta_t-\theta_0+\tilde{\sigma}))}{
    \Gamma(1-\tfrac{1}{2}(\tilde{\sigma}+\theta_\infty))
     \Gamma(1+\tfrac{1}{2}(\theta_t+\theta_0-\tilde{\sigma}))
     \Gamma(1+\tfrac{1}{2}(\theta_t-\theta_0-\tilde{\sigma}))}s_V.
   \label{eq:kappa}
\end{equation}

As a comment, the diagonal matrices $D_0$ and $D_t$ represent the
ambiguity in diagonalizing $M_t$ and $M_0$, which is in turn tied to
the choice of normalization of the Frobenius basis $y_i(z;z_0)$ at each
point. Likewise, $C_\infty$ diagonalizes $M_\infty$ and $D$ represents
the ambiguity in the basis normalization at $\infty$. The parameter
$\kappa$ (or $s_V$) then has the interpretation of the relative
normalization between the system at $\infty$ and the system at $0,t$,
which is an isomonodromy invariant as can be checked from the
asymptotic analysis like that in \cite{Jimbo:1982aa} or
\cite{Andreev:2000aa}. Alternatively, one can relate the
$s_V=e^{i\eta}$ to the relative twist between the ``gluing'' 
of the 3-point Riemann-Hilbert problem with monodromies
$\{\theta_0,\theta_t,\sigma\}$ -- which is solved by hypergeometric
functions -- to the 2-point irregular Riemann-Hilbert
problem $\{-\sigma,\theta_{\infty},s_1,s_2\}$ -- solved by confluent
hypergometrics -- as was defined in \cite{Lisovyy:2018mnj}. 

\subsection{The accessory parameter for the confluent Heun equation}

Solving \eqref{eq:difflogtau} involves finding the root of the JMU
$\tau$ function and then using the value of this root to find
$c_{t_0}$ as the derivative of the logarithm of the shifted
function. Given the structure of \eqref{eq:nekrasov}, it is
interesting to write
\begin{equation}
  \tau(\rho;t)=C_{V}(\vec{\theta};\sigma)
  t^{\frac{1}{4}(\sigma^2-\theta_0^2-\theta_t^2)}
  e^{-\frac{1}{2}\theta_tt}\hat{\tau}(\rho;t)
\end{equation}
where $\hat{\tau}$ involves only the combinatorial expansion of the
irregular conformal blocks \eqref{eq:irrcb} and ratios of Barnes
functions which can be written in terms of Euler's gamma
functions. The asymptotics of $\hat{\tau}$ is given by 
\cite{Jimbo:1982aa}:
\begin{multline}
  \hat{\tau}(\rho;t)=1+\left(\frac{\theta_t}{2}-\frac{\theta_\infty}{4}
    +\frac{\theta_\infty(\theta_0^2-\theta_t^2)}{4\tilde{\sigma}^2}\right)t
  \\ +
  \frac{(\theta_\infty-\tilde{\sigma})((\tilde{\sigma}+\theta_t)^2-
    \theta_0^2)}{8\tilde{\sigma}^2(\tilde{\sigma}-1)^2}\kappa^{-1}
  t^{1-\tilde{\sigma}}+
   \frac{(\theta_{\infty}+\tilde{\sigma})((\tilde{\sigma}-\theta_t)^2-
     \theta_0^2)}{8\tilde{\sigma}^2(\tilde{\sigma}+1)^2}
   \kappa\, t^{1+\tilde{\sigma}}+{\mathcal O}(t^2,t^{2\pm
    2\Re\tilde{\sigma}}),
  \label{eq:expansiontau}
\end{multline}
where parameter $\kappa$ is as above. The $\tilde{\sigma}$ appearing
in \eqref{eq:expansiontau} is related to the monodromy parameter by the
addition of an even integer $\tilde{\sigma}=\sigma-2p$,
$p\in\mathbb{Z}$. This indeterminacy stems from the 
quasi-periodicity of the Nekrasov expansion \eqref{eq:nekrasov} with
respect to $\sigma$:
\begin{equation}
  \tau(\vec{\theta},\sigma,s_V;t)=
  s_V^{-p}\tau(\vec{\theta},\sigma-2p,s_V;t), 
  \quad\quad p\in\mathbb{Z}.
\end{equation}
This quasi-periodicity will impose a multi-valuedness in the monodromy
parameters found by solving \eqref{eq:zerotau}. The non-trivial zeros
of $\tau$ are those of $\hat{\tau}$, but, to work the asymptotics we
have to make sure that the terms in the expansion
\eqref{eq:expansiontau} are indeed dominant. To that end, it is useful to
define the variable $\tilde{\kappa}=\kappa t^{\sigma}$. Seen as a
function of $\tilde{\kappa}$  and $t$, $\hat{\tau}$ is meromorphic in
$\tilde{\kappa}$ and so 
$\hat{\tau}(\tilde{\kappa},t_0)=0$ can be inverted to give
$\tilde{\kappa}(\vec{\theta},\sigma;t_0)$. The quasi-periodicity means
that, from one such solution, we can create a series labelled by the
integer $p$: 
\begin{equation}
  s_V(\vec{\theta},\sigma;t_0;p)=
    Y(\vec{\theta},\sigma-2p)
  t_0^{1-\sigma+2p}    X(\vec{\theta},\sigma-2p;t_0), 
  \quad\quad p\in\mathbb{Z},
  \label{eq:exseries}
\end{equation}
where $Y(\vec{\theta},\sigma)$ is related to the string of gamma
functions in \eqref{eq:kappa},
\begin{equation}
  Y(\vec{\theta},\sigma)=\frac{
     \Gamma^2(\sigma)
    \Gamma(\tfrac{1}{2}(2-\sigma-\theta_\infty))
    \Gamma(\tfrac{1}{2}(2-\sigma+\theta_{t}+\theta_0))
    \Gamma(\tfrac{1}{2}(2-\sigma+\theta_{t}-\theta_0))}{
    \Gamma^2(2-\sigma)
    \Gamma(\tfrac{1}{2}(\sigma-\theta_\infty))
     \Gamma(\tfrac{1}{2}(\sigma+\theta_{t}+\theta_0))
     \Gamma(\tfrac{1}{2}(\sigma+\theta_{t}-\theta_0))},
\end{equation}
and $X(\vec{\theta},\sigma;t_0)$ is analytic, obtained by inverting
\eqref{eq:expansiontau}. We quote the first three terms, valid if $\Re
\sigma>0$:
\begin{subequations}
\begin{equation}
  X(\vec{\theta},\sigma;t_0)
  = 1+\chi_1 t_0 +\chi_2
    t_0^2+\ldots + \chi_n t_0^n+\ldots
  \label{eq:ex}
\end{equation}
with
\begin{equation}
    \chi_1=(\sigma-1)\frac{\theta_\infty
      (\theta_0^2-\theta_{t}^2)}{\sigma^2(\sigma-2)^2},
  \end{equation}
and
  \begin{multline}
    \chi_2=\frac{\theta_\infty^2(\theta_0^2-\theta_{t}^2)^2}{64}
    \left(\frac{5}{\sigma^4}-\frac{1}{(\sigma-2)^4}
      -\frac{2}{(\sigma-2)^2}+\frac{2}{\sigma(\sigma-2)}\right)
    \\
  -\frac{(\theta_0^2-\theta_{t}^2)^2+2\theta_\infty^2
    (\theta_0^2+\theta_{t}^2)}{64}\left(
    \frac{1}{\sigma^2}-\frac{1}{(\sigma-2)^2}\right)
  \\
  +\frac{(1-\theta_\infty^2)((\theta_{0}-1)^2-\theta_{t}^2)((\theta_{0}+1)^2-
    \theta_{t}^2)}{128}\left(\frac{1}{(\sigma+1)^2}-\frac{1}{(\sigma-3)^2}\right).
  \end{multline}
\end{subequations}
The value of $p$ in \eqref{eq:exseries} will be determined, later, by
the requirement that the quantities have a sensible limit as
$t_0\rightarrow 0$. For the accessory parameter \eqref{eq:difflogtau},
this ambiguity is just the shift on $\sigma$ by an even integer, which
will play no further role.  In order to use \eqref{eq:difflogtau}
and find the accessory parameter, we must shift the monodromy
parameters by one unit. A simple calculation using \eqref{eq:kappa}
yields: 
\begin{equation}
  \tilde{\kappa}(\rho^-;t)=\frac{8\sigma^2(\sigma-1)^2}{ (\sigma-\theta_\infty)
    ((\sigma+\theta_t)^2-\theta_0^2)t}\tilde{\kappa}(\rho;t).
\end{equation}
Now, using \eqref{eq:difflogtau}
\begin{equation}
  c_{t_0}=\frac{(\sigma-1)^2-(\theta_0+\theta_{t}-1)^2}{4t_0}
  -\frac{\theta_{t}-1}{2}+\frac{d}{dt}\log\hat{\tau}(\rho^-;t_0),
\end{equation}
and expanding the $\hat{\tau}$ term, we find the asymptotic formula
for the accessory parameter
\begin{subequations}
\begin{equation}
  t_0c_{t_0}=k_0+k_1t_0+k_2t_0^2+\ldots +k_nt_0^n+\ldots,
  \label{eq:accessoryc}
\end{equation}
with the three first terms in the expansion given by
  \begin{equation}  
    k_0=\frac{(\sigma-1)^2-(\theta_0+\theta_{t}-1)^2}{4},\quad\quad
    k_1=-\frac{\theta_\infty(\sigma(\sigma-2)-\theta_0^2+\theta_{t}^2)}{
      4\sigma(\sigma-2)},
    \label{eq:accessoryc01}
  \end{equation}
  \begin{multline}
    k_2=\frac{1}{32}+\frac{\theta_\infty^2(\theta_0^2-\theta_{t}^2)^2}{64}
      \left(\frac{1}{\sigma^3}-\frac{1}{(\sigma-2)^3}\right)
      +\frac{(1-\theta_\infty^2)(\theta_0^2-\theta_{t}^2)^2+2\theta_\infty^2
        (\theta_0^2+\theta_{t}^2)}{32\sigma(\sigma-2)}
      \\-
        \frac{(1-\theta_\infty^2)((\theta_0-1)^2-\theta_{t}^2)((\theta_0+1)^2-
          \theta_{t}^2)}{32(\sigma+1)(\sigma-3)},
     \label{eq:accessoryc2}
  \end{multline}
\end{subequations}
where we assumed $\Re \sigma>0$. The corresponding expression for $\Re
\sigma<0$ can be obtained by sending $\sigma\rightarrow
-\sigma$. Higher order terms can be consistently computed using 
\eqref{eq:nekrasov}. Although the terms become increasingly
complicated, we have the structure where
the term $k_n$ is a rational function of the monodromy parameters,
and analytic in the single monodromy parameters $\vec{\theta}$. As a
function of $\sigma$ it is meromorphic, with poles at integer
values. The structure of the poles at order $n$ is
\begin{itemize}
  \item poles of order $2n-1$ and below at $\sigma=0$ and $\sigma=\pm 2$;
  \item single poles at $\sigma=\pm 3,\ldots,\pm(n+1)$ -- note
    that the structure of \eqref{eq:accessoryc} for negative $\sigma$
    is illusory, since it is only valid for $\Re\sigma >0$;
 \item analytic at $\sigma=1$.
\end{itemize}
This structure mirrors that of the accessory parameter for the
(non-confluent) Heun equation found in
\cite{Barragan-Amado:2018pxh}. There, the structure was inherited from
the corresponding structure of conformal blocks
\cite{Zamolodchikov:1987aa}. It seems that irregular conformal blocks 
display the same traits.

It should be stressed that \eqref{eq:zerotau} and
\eqref{eq:difflogtau} are exact relations, even though their
usefulness stems from our ability to compute the $\tau$ function for
Painlevé V efficiently. Miwa's theorem \cite{Miwa:1981aa} shows
that the $\tau$ function is analytic in the whole complex plane except
at $t=0$ and $t=\infty$. Thus, the expansion \eqref{eq:nekrasov} has
infinite radius of convergence, even if it becomes exponentially hard
to compute the higher order coefficients in $t$, due to their
combinatorial nature. These limitations should be overcome by the
Fredholm determinant formulation of the $\tau$ function proposed
recently \cite{Lisovyy:2018mnj}, which would be of great help for
numerical studies.

At $t=\infty$, the expansion of the Painlevé V $\tau$ function is
substantially more complicated. No general expansion exists, but
formulas for $t\rightarrow \infty$ along specific rays, such as
$\arg t = 0,\pi/2,\pi,3\pi/2$ have been proposed, see
\cite{Lisovyy:2018mnj} for a review as well as the relation between
these expansions and the different types of irregular conformal blocks
at $c=1$. In the application of interest in this work, however, the
parameter $t_0$ depends on $\omega$, which will be complex for the
general case, therefore straying from these rays. We hope to study the
large frequency asymptotic of the quasi-normal modes in the context
presented here in future work.

\section{Spheroidal Harmonics}
\label{sec:angular}

We are interested in solutions of \eqref{eq:confluentheun} which are
regular at both the South and the North poles:
\begin{equation}
  y(z)=\begin{cases}
    z^{0}(1+{\mathcal O}(z)), & z\rightarrow 0; \\
    (z-t_0)^{0}(1+{\mathcal O}(z-1)), & z\rightarrow t_0;
  \end{cases}
  \label{eq:boundary}
\end{equation}
which will place a restriction on the value of $\lambda$, allowing
only a discrete set as possible values $\lambda_\ell(s,m)$, $\ell\in
\mathbb{N}$. Finding these correspond to the \textit{eigenvalue
  problem} for the angular equation. 

We are going to define the single monodromy parameters
\begin{equation}
  \theta_{0}=-m-s, \quad\quad \theta_{t_{0}}=m-s,
  \quad\quad \theta_{\infty}=2s.
  \label{eq:singlemonoangular}
\end{equation}
Upon the change of variables
\begin{equation}
  y(z) =
  (1+\cos\theta)^{\theta_{t_{0}}/2}(1-\cos\theta)^{\theta_{0}/2}
  S(\theta),\quad\quad
  z = -2a\omega(1-\cos\theta),
  \label{eq:shomotopic}
\end{equation}
we bring the differential equation to a canonical confluent Heun form
\eqref{eq:confluentheun}, with $\vec{\theta}$ as above and
\begin{equation}
  t_{0}= -4a\omega, \quad\quad  t_{0}c_{t_{0}} =\lambda+2a\omega
  s+a^2\omega^2.
  \label{eq:accessoryangular}
\end{equation}
Given the expansion \eqref{eq:accessoryc}, it is a matter of direct
substitution of the parameters of the spheroidal harmonic equation
\eqref{eq:singlemonoangular} and \eqref{eq:accessoryangular}, using
the quantization condition \eqref{eq:quantization}:
\begin{equation}
  \theta_0=-m-s,\quad\quad \theta_{t_0}=m-s,\quad\quad
  \theta_\infty = 2s,\quad\quad t_0=-4a\omega,\quad\quad
  \sigma = -2s+2j.
\end{equation}
The result is:
\begin{multline}
  {_s\lambda}_{\ell,m}(a\omega)=
  (\ell-s)(\ell+s+1)-\frac{2ms^2}{\ell(\ell+1)}a\omega \\
  +\left(\frac{2((\ell+1)^2-m^2)((\ell+1)^2-s^2)^2}{(2\ell+1)(\ell+1)^3(2\ell+3)}
    -\frac{2(\ell^2-m^2)(\ell^2-s^2)^2}{(2\ell-1)\ell^3(2\ell+1)}
    -1\right)a^2\omega^2+{\mathcal O}(a^3\omega^3),
  \label{eq:angulareigenvalue}
\end{multline}
which can be checked to agree with the literature \cite{Seidel:1988ue}
-- see \cite{Berti:2005gp} for a thorough review. In order to
recover the asymptotics, we chose $j=\ell+s+1$ in
\eqref{eq:quantization}. As anticipated in \cite{Teukolsky:1973ha},
the minimum eigenvalue of $\ell$ is $|s|$ and the azimuthal momentum
is constrained so $|m|\leq\ell$.

\section{Conformal blocks and the radial equation}
\label{sec:radial}

The Nekrasov expansion for some of the Painlevé $\tau$ functions has
been interpreted in terms of $c=1$ conformal blocks in
\cite{Gamayun:2012ma,Gamayun:2013auu}.  The details of the structure
stems from the AGT conjecture \cite{Alday:2009aq,Alba:2010qc} and can
be checked in the references. For Painlevé VI, the structure of the
corresponding $\tau$ function is similar to \eqref{eq:nekrasov}, with
``instanton sectors'' labelled by $n$, and regular conformal blocks,
defined as 
\begin{equation}
  {\mathcal F}\left(\substack{P_{1}\;\quad 
      P_{t}\\ P \\ P_{\infty}\quad P_0};t\right)
   = \langle \Delta_{\infty}|V_{\Delta_1}(1)\Pi_{\Delta}
   V_{\Delta_t}(t)|\Delta_{0}\rangle,\quad\quad
   \Delta_k = \frac{c-1}{24}+P_k^2,\quad
   \Delta = \frac{c-1}{24}+P^2,
\end{equation}
where $V_{\Delta_i}(z_i)$ are primary vertex
operators, acting on the primary state $|\Delta_j\rangle$ and its
descendants (the Verma module built on the primary state) with an
operator of dimension $\Delta_i$ 
and $\Pi_\Delta$ a projector onto the Verma module generated from
$|\Delta\rangle$ (see \cite{Ribault:2014hia} for details and
notation). The conformal blocks are dependent on the Virasoro Algebra
central charge $c$ -- which enters through the Kac-Shapovalov matrix of
inner products of descendant states of $|\Delta\rangle$. ${\mathcal
  F}$ can be seen to have the asymptotic expansion
\begin{equation}
  {\mathcal F}\left(\substack{P_{1}\;\quad 
       P_{t}\\  P \\ P_{\infty}\quad P_0};t\right)
   =
   t^{\Delta-\Delta_t-\Delta_0}(1-t)^{-2(\tfrac{Q}{2}-iP_t)\tfrac{Q}{2}-iP_1)}\left(1
     +\frac{(\Delta-\Delta_1+\Delta_2)(\Delta-\Delta_4+\Delta_3)}{2\Delta}t+
     \mathcal O(t^2)\right),
\end{equation}
where again $c=1+6Q^2$. The higher order terms in $t$ can be computed either
recursively \cite{Zamolodchikov:1987aa} or via the Nekrasov functions
\cite{Nekrasov:2002qd,Alba:2010qc}.  

The correspondence between $c=1$ conformal blocks and accessory
parameters to Fuchsian ordinary differential equations have been
stablished in \cite{Novaes:2014lha}. On the other hand, in
\cite{Litvinov:2013sxa} it was outlined a method to compute the
same accessory parameters using \textit{semi-classical blocks}. These
are obtained in the $c\rightarrow \infty$ limit of the conformal
blocks defined above. The analysis at the semi-classical limit is
based on the property of exponentiation:
\begin{equation}
  {\mathcal F}\left(\substack{P_{1}\;\quad 
       P_{t}\\ P \\ P_{\infty}\quad P_0};t\right)
   \simeq\exp\left[\frac{c}{6}{\mathscr{F}}(\delta_k;\delta;t)\right]
\end{equation}
where $\delta_k,\delta$ are obtained from a scaling procedure from
$\Delta_k,\Delta$. With the parametrization $Q=b+1/b$, we have
\begin{equation}
  \delta_k=\lim_{b\rightarrow 0}b^2\Delta_k,\quad\quad
  \delta = \lim_{b\rightarrow 0}b^2\Delta.
\end{equation}

It can be checked by applying the Virasoro algebra that the Verma
module constructed from the ``light'' operator $V_{(2,1)}(z)$, with
$\Delta_{(2,1)}=-\frac{1}{2}-\frac{3b^2}{4}$ has a null vector at level
2. Requiring that this vector decouples from correlation functions
imply the condition
\begin{equation}
  \frac{1}{b^2}\frac{\partial^2}{\partial z_0^2} V_{(2,1)}(z_0) +:T(z)V_{(2,1)}(z_0):=0.
\end{equation}
When this condition is applied to correlation functions involving
primary operators, we find Fuchsian differential equations,
essentially due to the fact that the OPE between $T(z)$ and primary
operators at $z_i$ have no terms diverging faster than $(z-z_i)^{-2}$. 

To describe irregular singular points we need to take confluent limits
of two primary operators \cite{Gaiotto:2009ma}, which are associated
to Whittaker modules of the Virasoro algebra, see
\cite{Nagoya:2015cja} for a review and \cite{Lisovyy:2018mnj} for the
relation between these conformal blocks to the asymptotics of Painlevé V.
The confluent limit of the two colliding primary operators generates
an non-primary operator, and in order to derive the corresponding Ward
identity related to the null condition we will work with the
Feigin-Fuchs representation of Liouville field theory,
\begin{gather}
  \phi(z_0)\phi(z_1)=-\frac{1}{2}\log|z_0-z_1|^2+:\phi(z_0)\phi(z_1):\quad\quad
  T(z)=-:(\partial\phi)^2(z):+Q\partial^2\phi(z),\nonumber \\
  \Delta(:e^{2\alpha\phi(z)}:)=\alpha(Q-\alpha),
  \label{eq:liouvilledefs}
\end{gather}
which can be seen to generate a central charge of $c=1+6Q^2$. The
confluent primary vertex operator of rank $1$, as defined in 
\cite{Nagoya:2010yb}, is given by
\begin{equation}
  V_{\alpha,\beta}(z)=\,\,:\exp\left(2\alpha\phi(z)+2\beta\partial\phi(z)\right):,
\end{equation}
and by analogy with primary operators and Verma modules,
$V_{\alpha,\beta}(0)$ is associated to an Whittaker module of states.
We find for singular terms of the OPE with the stress-energy tensor
\begin{multline}
  T(z)V_{\alpha,\beta}(z_i)=-\left(\frac{\alpha}{z-z_i}+
    \frac{\beta}{(z-z_i)^2}\right)^2V_{\alpha,\beta}(z_i)+
  \frac{1}{z-z_i}\frac{\partial}{\partial z_i} V_{\alpha,\beta}(z_i) \\
  + 
  \frac{1}{(z-z_i)^2}\beta\frac{\partial}{\partial\beta}
  V_{\alpha,\beta}(z_i)+
  Q\left(\frac{\alpha}{(z-z_i)^2}+\frac{2\beta}{(z-z_i)^3}\right)
  V_{\alpha,\beta}(z_i)+\text{reg.},
\end{multline}
and then
\begin{equation}
  [L_n,V_{\alpha,\beta}(z_i)]=\left[
    z_i^{n+1}\frac{\partial}{\partial z_i}+(n+1)z_i^n\left(
      \Delta+\beta\frac{\partial}{\partial\beta}\right)
    +n(n+1)z_i^{n-1}\beta(Q-\alpha)+
    \frac{n(n^2-1)}{6}z_i^{n-2}\beta^2\right]V_{\alpha,\beta}(z_i),
\end{equation}
where $\Delta=\alpha(Q-\alpha)$.
The global conformal Ward identities on the correlation functions of
$N$ Whittaker operators follow from the commutation relations
$[L_n,V_{\alpha,\beta}(z_i)]$ for $n=-1,0,1$:
\begin{equation}
  \sum_{i=1}^N\frac{\partial}{\partial z_i}=0,\quad\quad
  \sum_{i=1}^Nz_i\frac{\partial}{\partial z_i}+\Delta_i+\beta_i
  \frac{\partial}{\partial \beta_i}=0,\quad\quad
    \sum_{i=1}^Nz^2_i\frac{\partial}{\partial z_i}+2z_i\Delta_i+2z_i\beta_i
  \frac{\partial}{\partial \beta_i}+2\beta_i(Q-\alpha_i)=0.
\end{equation}
Note that these expressions reduce to the well-known formulas
involving primary operators if we take $\beta_i=0$.

For the confluent Heun equation \eqref{eq:confluentheun}, the relevant
conformal block has 3 insertions of primary operators and one with a
non-trivial Whittaker operator: 
\begin{equation}
  \langle V_{(2,1)}(z_0) V_{\alpha_1}(z_1)V_{\alpha_2}(z_2)\Pi_\Delta
  V_{\alpha_3,\beta}(z_3)\rangle =
  \mathscr{G}_{b}(z_i;\alpha_i,\Delta,\beta),
  \label{eq:ccb}
\end{equation}
using the global conformal Ward identities to solve for
$\partial_{z_i}\mathscr{G}_b$ and setting $z_3=0$, $z_2=1$ and
$z_1=\infty$, we find that the null vector condition is
\begin{multline}
  \left(\frac{1}{b^2}\frac{\partial^2}{\partial z_0^2}-
    \left(\frac{1}{z}+\frac{1}{z-1}\right)\frac{\partial}{\partial
      z_0}+\frac{\Delta_1}{z_0^2}+\frac{\Delta_2}{(z_0-1)^2}\right.\\
  \left. +
    \frac{\Delta_3-\Delta_{(2,1)}-\Delta_1-\Delta_2}{z_0(z_0-1)}
    +\frac{2\beta(Q-\alpha)}{z_0^3}-\frac{\beta^2}{z_0^4}
  +\frac{1}{z_0(z_0-1)}\beta\frac{\partial}{\partial
 \beta}\right)\mathscr{G}_b=0,
\end{multline}
The semiclassical limit is obtained
through the scaling:
\begin{equation}
  \alpha_i=\frac{\eta_i}{b},\quad\quad
  \beta = \frac{t_0}{2b},\quad\quad
  \Delta = \frac{\delta}{b^2}.
\end{equation}
In this limit, the three insertions at $0,1,\infty$ become ``heavy'',
and set the background over which the ``light'' operator
$V_{(2,1)}(z_0)$ will induce fluctuations.  Assuming exponentiation, the
four-point function \eqref{eq:ccb} should factorize as
\begin{equation}
  \langle V_{(2,1)}(z_0) V_{\alpha_1}(\infty)V_{\alpha_2}(1)\Pi_\Delta
  V_{\alpha_3,\beta}(0)\rangle\bigg\rvert_{b\rightarrow 0} \simeq
  \psi(z_0;\delta_i,\delta;t_0)
  \exp\left(\frac{1}{b^2}\mathscr{B}(\delta_i,\delta;t_0)\right),
\end{equation}
where $\mathscr{B}$ is the semi-classical confluent conformal block of
the first kind, defined by analogy with $\mathscr{F}$
above. Setting $z_0=t_0/z$, we have for $\tilde{\psi}=z\psi$, as
$b\rightarrow 0$, 
\begin{equation}
  \frac{\partial^2}{\partial z^2}\tilde{\psi}+\left(
    -\frac{1}{4}+\frac{\delta_1}{z^2}+\frac{\delta_2}{(z-t_0)^2}
    +\frac{1-\eta_3}{z}
    +\frac{\delta-\delta_2-\delta_1}{z(z-t_0)}\right)\tilde{\psi}=0.
  \label{eq:cccb}
\end{equation}
with
\begin{equation}
  \delta_i=\eta_i(1-\eta_i),\,\, i=1,2,3;\quad\quad\quad\quad 
  \delta=\eta_3(1-\eta_3)+t_0\frac{\partial}{\partial t_0}
  \mathscr{B}.
\end{equation}
The last equality for $\delta$ is required by the projection operator
$\Pi_\Delta$. 

The equation \eqref{eq:cccb} can be cast in the canonical confluent
Heun form \eqref{eq:confluentheun} if we write it in terms of
\begin{equation}
  y(z)=z^{-\eta_1}(z-t_0)^{-\eta_2}\tilde{\psi}(z)
\end{equation}
and set
\begin{gather}
  \eta_1=\frac{1-\theta_0}{2},\quad\quad
  \eta_2=\frac{1-\theta_{t_0}}{2},\quad\quad
  \eta_3=1-\frac{\theta_{\infty}}{2},\\
  t_0c_{t_0}=-t_0\frac{\partial \mathscr{B}}{\partial t_0}
  -(\eta_1+\eta_2+\eta_3-1)(\eta_1+\eta_2-\eta_3)
  =-t_0\frac{\partial \mathscr{B}}{\partial
    t_0}+\frac{(\theta_\infty-1)^2}{4}
  -\frac{(\theta_0+\theta_{t_0}-1)^2}{4}.
\end{gather}
Comparing with the expression \eqref{eq:accessoryc}
above, we can compute $\mathscr{B}$:
\begin{equation}
  \mathscr{B}\left(\theta_\infty;\sigma;
    \substack{\theta_{t} \\ \theta_{0}};t_0\right)
  =\frac{(\theta_\infty-1)^2-(\sigma-1)^2}{4}\log t_0-
  k_1t_0-\frac{k_2}{2}t_0^2-\ldots -\frac{k_{n}}{n}t_0^{n}+\ldots,
  \label{eq:bcequalsone}
\end{equation}
with $k_n$ given as \eqref{eq:accessoryc01} and
\eqref{eq:accessoryc2}. 
This expression for the irregular classical conformal block can be
confronted with the $b\rightarrow 0$ limit of the confluent conformal
block \eqref{eq:confluentconformal}:
\begin{equation}
  \tilde{\mathscr{B}}(\delta_i,\delta;t) = \lim_{b\rightarrow 0}b^2\log
  \mathcal{B}\left(P_\infty;P; 
    \substack{P_{t} \\  P_{0}}; t/b\right),\quad\quad\quad\quad
  P_i=\frac{\theta_i}{2b}, \quad P=\frac{\sigma-1}{2b},
  \label{eq:bscalinglimit}
\end{equation}
which yields
\begin{equation}
  \tilde{\mathscr{B}}(\delta_i,\delta;t)=
  \frac{(\sigma-1)^2-\theta^2_0-\theta^2_t+1}{4}
  \log t - k_1t - \frac{k_2}{2}t-\ldots
  - \frac{k_n}{n}t^n+\ldots,
  \label{eq:bsemiclassical}
\end{equation}
with the difference in the leading term stemming from the different
normalization conditions on the Whittaker vector
$V_{\alpha,\beta}(0)$. The expansion \eqref{eq:confluentconformal}
requires $\langle \Delta | \alpha,\beta \rangle = 1$ in the
semiclassical limit, whereas \label{eq:bcequalsone} is computed
entirely from $c=1$ blocks.

\subsection{The radial equation}

After this long exposition we can turn to the radial equation \eqref{eq:radialeq}.
The Teukolsky master equation for the radial part is
\begin{equation}
\Delta^{-s}\frac{d}{dr}\left(\Delta^{s+1}\frac{dR(r)}{dr}\right)+
\left(\frac{K^{2}(r)-2is(r-M)K(r)}{\Delta}+4is\omega
  r-{_s\lambda_{\ell,m}}-a^2\omega^2+2am\omega
\right)R(r)=0,
\end{equation}
where
\begin{equation}
  K(r)=(r^2+a^2)\omega-am,\quad\quad \Delta =
  r^2-2Mr+a^2=(r-r_+)(r-r_-).
\end{equation}
In order to bring it to our canonical form, let us define
\begin{gather}
  \theta_{-}= s - i \frac{\omega-m\Omega_{-}}{2\pi T_-}, \quad\quad
  \theta_{+}=  s + i \frac{\omega-m\Omega_{+}}{2\pi T_+},\quad\quad
  \theta_{\infty}=2s-4iM\omega,\\
  2\pi T_{\pm} = \frac{r_+-r_-}{4Mr_{\pm}}, \quad\quad
  \Omega_{\pm} = \frac{a}{2Mr_{\pm}}.
  \label{eq:singlemonos}
\end{gather}
By changing variables
\begin{equation}
  R(r)
  =(r-r_{-})^{-(\theta_{-}+s)/2}(r-r_{+})^{-(\theta_{+}+s)/2}y(r),
  \quad\quad z= 2i\omega(r-r_{-}),
\end{equation}
we arrive at
\begin{equation}
  \frac{d^{2}y}{dz^{2}}+\left[\frac{1-\theta_{-}}{z}+\frac{1-\theta_{+}}{z-z_{0}}
  \right] \frac{dy}{dz}+
  \left[-\frac{1}{4}+\frac{\theta_\infty}{2z}-\frac{z_0c_{z_{0}}}{z(z-z_{0})}
  \right]y(z)=0,
\end{equation}
with
\begin{equation}
  z_0=2i(r_+-r_-)\omega,\quad\quad
   z_{0}c_{z_{0}} =
   {_s\lambda}_{\ell,m}+2s+2i(1-2s)M\omega-is(r_+-r_-)\omega-
   (2r_++r_-)r_+\omega^2.
\end{equation}
We note the following relations
\begin{equation}
  \theta_-+\theta_{+}+\theta_\infty = 4s,\quad\quad
  \theta_-+\theta_{+}-\theta_\infty =4i(r_++r_-)\omega,
  \label{eq:radialrelations}
\end{equation}

The relation between the accessory parameter and the semiclassical
confluent conformal block allow us to interpret the single monodromy
parameters $\theta_i$ as Liouville momenta. The Regge-Okamoto symmetry
of the confluent conformal block \cite{Lisovyy:2018mnj}
\begin{equation}
  \mathcal{B}\left(P_\infty; P;
    \substack{P_{t} \\ P_{0}};t\right)=
  t^{\frac{1}{2}(P_\infty^2-(P_0+P_t)^2)}e^{\frac{1}{2}\Pi t}
  \mathcal{B}\left(P_\infty-2\Pi;P;
    \substack{P_{t}-\Pi \\ P_{0}-\Pi};t\right),\quad\quad
  \Pi = \tfrac{1}{2}(P_0+P_t+P_\infty),
\end{equation}
when applied to the radial equation \eqref{eq:singlemonos}, due to the
relations \eqref{eq:radialrelations}, allow for
the following association to the Liouville momenta of the primary
insertions at $z=0,t_0$. From the assignment \eqref{eq:bscalinglimit}
and the relation \eqref{eq:radialrelations}, we have $\Pi = s/b$ and
the shifted momenta  
\begin{equation}
  P_+\equiv P_{t}-\Pi =
  \frac{i}{4\pi b}\frac{\omega - m\Omega_+}{T_+},\quad\quad
  P_-\equiv P_0-\Pi=-\frac{i}{4\pi b}\frac{\omega-m\Omega_-}{T_-},\quad\quad
  P_\infty-2\Pi = -\frac{4iM\omega}{b},
  \label{eq:liouvillemomenta}
\end{equation}
which, just as the analogue in the scalar case in five dimensions
\cite{Amado:2017kao}, has the interpretation of entropy influx at the
horizons given a quanta of energy $\omega$ and angular momentum
$m$. If we take $b$ to be purely imaginary, these are real numbers for
real $\omega$. Note that these expressions make sense only in the
$b\rightarrow 0$ limit.

In order to phrase the quantization condition for the radial equation
in terms of monodromy data, In terms of $y(z)$ the radial boundary
conditions of purely ingoing wave at infinity ($r\rightarrow \infty$)
and purely outgoing at the horizon ($r\rightarrow r_+$) are
\cite{Leaver:1985ax}, 
\begin{equation}
  y(z)=\begin{cases}
    e^{\tfrac{1}{2}z}(1+{\mathcal O}(z^{-1})), & z\rightarrow +i\infty; \\
    1+{\mathcal O}(z-z_0), & z\rightarrow z_0,
  \end{cases}
  \label{eq:boundaryradial}
\end{equation}
the relative normalizations can be worked out but are not relevant to
this problem. We note that the first condition is rather at
$z\rightarrow +i\infty$ due to the relation between it and the radial
coordinate $z=2i\omega(r_+-r_-)$, assuming $\Re \omega >0$. Because of
the Stokes phenomenon, we can only guarantee that the solution of the
radial equation will display the behavior \eqref{eq:boundaryradial} in
a sector containing the ray $\arg z=\pi/2$.

We see that the solutions from the first row of the
fundamental matrix \eqref{eq:fundamentalmatrix} of the matrix system
$y_1(z)$ and $y_2(z)$ satisfying \eqref{eq:asympphik} correspond to
(non-normalized) Jost functions for the scattering problem: purely
ingoing waves at $z=+i\infty$ and outgoing at $z=z_0$. We need some
care to translate this to conditions on the connection matrices,
because the usual parametrization for the monodromy matrices, given in
Sec. \ref{sec:monomatrices}, assumes that $M_\infty$ is diagonal. The conditions
\eqref{eq:boundaryradial} asks that one compares the Frobenius basis
at $z=t$ -- in which the monodromy matrix $M_t$ is diagonal, with the
``Jost'' basis at infinity where the boundary conditions for the
fundamental matrix are given by \eqref{eq:asympphik}. This basis is
sometimes called ``Floquet'', or path-multiplicative basis
\cite{Ronveaux:1995}. In this basis, the monodromy matrix at infinity
is given by \eqref{eq:monoinfty}. 
 
In order to show that this basis do not change under the isomonodromy
flow, consider the Schlesinger equations \eqref{eq:schlesinger}
equations. By them, the quantities
\begin{equation}
  \Tr(A_0+A_t)=\hat{\theta}_0+\hat{\theta}_t,\quad\quad
  \Tr[\sigma_3(A_0+A_t)]= -\hat{\theta}_\infty
\end{equation}
are isomonodromy invariantes. Therefore, the diagonal elements of
$A_0+A_t$ are invariant under the flow. These elements set the
asymptotic form of the solution at $z=\infty$ (in $\mathcal{S}_2$) to
be \eqref{eq:asympphik}. Write
\begin{equation}
  \Phi_k(z)=\left(\mathbbold{1}+\frac{B_1}{z}+\ldots\right)
  z^{B_0}e^{\frac{1}{2}\sigma_3z}
\end{equation}
for the solution of \eqref{eq:matrixsystem} near $z=\infty$. The
existence of this limit requires that the matrix $B_0$ is
diagonal. The subleading term gives
\begin{equation}
  B_0+\frac{1}{2}[B_1,\sigma_3]=A_0+A_t
\end{equation}
Since the diagonal terms of $[B_1,\sigma_3]$ vanish, the off-diagonal
elements of $A_0+A_t$ only alter the subleading $\mathcal{O}(1/z)$
terms of $\Phi_k(z)$, and then preserve the asymptotic form of the
wavefunction.

The condition that the connection matrix between $z=+i\infty$ and
$z=z_0$ is lower triangular can be read from the explicit
representation \eqref{eq:connectioninfty},
\begin{equation}
  \kappa = \frac{\Gamma^2(1-\sigma)}{\Gamma^2(1+\sigma)}
  \frac{\Gamma(1+\frac{1}{2}(\sigma-\theta_\infty))
    \Gamma(-\frac{1}{2}(\theta_++\theta_--\sigma))
    \Gamma(-\frac{1}{2}(\theta_+-\theta_--\sigma))}{
    \Gamma(1-\frac{1}{2}(\sigma-\theta_\infty))
    \Gamma(-\frac{1}{2}(\theta_++\theta_-+\sigma))
    \Gamma(-\frac{1}{2}(\theta_+-\theta_-+\sigma))}.
\end{equation}
Comparing to the analogue problem of finding quasinormal modes in
Kerr-AdS$_5$ black holes \cite{Barragan-Amado:2018pxh}, the usefulness
of this expression is somewhat wanting. The lack of natural small
parameters makes it difficult to study radial eigenmodes
analytically. They can, however, be studied numerically. The
isomonodromy method will most likely not be as fast as Leaver's method
\cite{Leaver:1985ax}, but on the other hand we have more control on
the analyticity of the functions involved. The investigation is under
way and will be reported elsewhere.

\section{Discussion}
\label{sec:discussion}

In this work we considered the Teukolsky master equation eigenvalue
problem tackled by the isomonodromy method. We have seen from a more
general perspective the relationship between the ensuing confluent
Heun equations and confluent conformal blocks, built on Whittaker
modules. The mixture of classical complex analysis, integrable systems
(through Riemann-Hilbert problems) and conformal blocks has been drawn
some attention of late \cite{Dubrovin:2018aa,Dunne:2019aqp}, and we
have found in this paper that the eigenvalue problems for both the
angular and radial equation can be cast in terms of monodromy data and
solved by expansions of the Painlevé V $\tau$ function.

Using the Painlevé V small isomonodromic time expansion
\cite{Gamayun:2013auu,Lisovyy:2018mnj}, we derived expansions for the
spheroidal harmonic angular eigenvalue in terms of the frequency. We
have verified heuristically the exponentiation property for
semiclassical confluent conformal blocks (of the first type as defined
in \cite{Gaiotto:2009ma}) and used then to rederive the small $t$
expansion of the composite monodromy parameter $\sigma$. In turn, this
allowed us to interpret the radial equation as the null condition of a
composition of two primary operators at the radial positions of the
inner and outer horizon and a Whittaker operator seated at radial
infinity. Curiously, the primary operators can be seen to have real
Liouville momenta, in an ``unitary'' description of sorts.

Using the relation between the accessory parameter and the zero of the
isomonodromic tau function, we have an effective way to compute the
monodromy parameters -- and thus the connection matrix -- of the
solutions of the confluent differential equation
\eqref{eq:confluentheun}. This gives an effective algorithm to compute
scattering data and solving the eigenvalue problem which is
procedural. We are currently investigating methods to efficiently
compute the quasinormal modes in the notoriously hard quasi-extremal
regime ($r_-\rightarrow r_+$). While in all probability the method
will not be as fast as existing numerical methods for computing
quasinormal modes -- see \cite{Berti:2009kk}, the analytical
properties of monodromy parameters make a precision study amenable, as
well as enhancements in precision. 

The basic ingredients involved in the analysis are the monodromy
parameters and their relation to primary/Whittaker operators of a
CFT. We have found for both the radial and angular equations that
these monodromy parameters are associated to CFTs which can be
considered unitary, and in the radial equation the Liouville momentum
of the operator at the outer horizon is proportional to the entropy
intake. The angular eigenvalue condition has again the interpretation
of an equilibrium condition between the ``angular'' and ``radial''
systems, just like the lore in \cite{Barragan-Amado:2018pxh}. It is
tempting to try to interpret \eqref{eq:liouvillemomenta} as small
perturbations on a ``macroscopic'' black hole state -- the CFT vacuum
in this case, and by composing these perturbations arrive at some
macroscopic state corresponding to a different black hole. Given the
$c=1$ interpretation of the process, we can perhaps count the
difference in the number of states using known facts about the
representation of Virasoro algebra. We leave these as enticing
prospects for future work. 

\section*{Acknowledgements}
The authors would like to thank Oleg Lisovyy for his special
assistance during the final part of the project. We are also indebted
to Julián Barragán-Amado, Fábio Novaes, Marc Casals, Jacopo Viti, 
Filiberto Aires, Dmitry Melnikov and Carlos Batista for discussions
and suggestions along the way. BCdC is also specially thankful for the
hospitality of ICTP-Trieste where this work was finished. This work
was partially funded by CAPES and CNPq.


\begin{thebibliography}{10}

\bibitem{Wald:1984}
R.~M. Wald, {\em {General Relativity}}.
\newblock The University of Chicago Press, 1984.

\bibitem{Abbott:2016blz}
{\bf LIGO Scientific, Virgo} Collaboration, B.~P. Abbott et~al., {\it
  {Observation of Gravitational Waves from a Binary Black Hole Merger}},  {\em
  Phys. Rev. Lett.} {\bf 116} (2016), no.~6 061102,
  [\href{http://arxiv.org/abs/1602.03837}{{\tt arXiv:1602.03837}}].

\bibitem{Akiyama:2019cqa}
{\bf Event Horizon Telescope} Collaboration, K.~Akiyama et~al., {\it {First M87
  Event Horizon Telescope Results. I. The Shadow of the Supermassive Black
  Hole}},  {\em Astrophys. J.} {\bf 875} (2019), no.~1 L1,
  [\href{http://arxiv.org/abs/1906.11238}{{\tt arXiv:1906.11238}}].

\bibitem{Teukolsky:1973ha}
S.~A. Teukolsky, {\it {Perturbations of a rotating black hole. I. Fundamental
  equations for gravitational, electromagnetic, and neutrino-field
  perturbations}},  {\em The Astrophysical Journal} {\bf 185} (1973) 635--648.

\bibitem{Chandrasekhar1983}
S.~Chandrasekhar, {\em {The mathematical theory of black holes}}, vol.~69.
\newblock Oxford University Press, 1983.

\bibitem{Berti:2005ys}
E.~Berti, V.~Cardoso, and C.~M. Will, {\it {On gravitational-wave spectroscopy
  of massive black holes with the space interferometer LISA}},  {\em Phys.
  Rev.} {\bf D73} (2006) 064030,
  [\href{http://arxiv.org/abs/gr-qc/0512160}{{\tt gr-qc/0512160}}].

\bibitem{Berti:2009kk}
E.~Berti, V.~Cardoso, and A.~O. Starinets, {\it {Quasinormal modes of black
  holes and black branes}},  {\em Class. Quant. Grav.} {\bf 26} (2009) 163001,
  [\href{http://arxiv.org/abs/0905.2975}{{\tt arXiv:0905.2975}}].

\bibitem{Berti:2005gp}
E.~Berti, V.~Cardoso, and M.~Casals, {\it {Eigenvalues and eigenfunctions of
  spin-weighted spheroidal harmonics in four and higher dimensions}},  {\em
  Phys.Rev.} {\bf D73} (2006) 024013,
  [\href{http://arxiv.org/abs/gr-qc/0511111}{{\tt gr-qc/0511111}}].

\bibitem{Casals:2009zh}
M.~Casals, S.~R. Dolan, A.~C. Ottewill, and B.~Wardell, {\it {Self-Force
  Calculations with Matched Expansions and Quasinormal Mode Sums}},  {\em Phys.
  Rev.} {\bf D79} (2009) 124043, [\href{http://arxiv.org/abs/0903.0395}{{\tt
  arXiv:0903.0395}}].

\bibitem{Motl:2003cd}
L.~Motl and A.~Neitzke, {\it {Asymptotic black hole quasinormal frequencies}},
  {\em Adv.Theor.Math.Phys.} {\bf 7} (2003) 307--330,
  [\href{http://arxiv.org/abs/hep-th/0301173}{{\tt hep-th/0301173}}].

\bibitem{Neitzke2007}
A.~Neitzke, {\it Greybody factors at large imaginary frequencies},
  \href{http://arxiv.org/abs/hep-th/0304080}{{\tt hep-th/0304080}}.

\bibitem{Castro2013}
A.~Castro, J.~M. Lapan, A.~Maloney, and M.~J. Rodriguez, {\it {Black Hole
  Monodromy and Conformal Field Theory}},  {\em Phys.Rev.} {\bf D88} (2013)
  044003, [\href{http://arxiv.org/abs/1303.0759}{{\tt arXiv:1303.0759}}].

\bibitem{Castro2013b}
A.~Castro, J.~M. Lapan, A.~Maloney, and M.~J. Rodriguez, {\it {Black Hole
  Scattering from Monodromy}},  {\em Class.Quant.Grav.} {\bf 30} (2013) 165005,
  [\href{http://arxiv.org/abs/1304.3781}{{\tt arXiv:1304.3781}}].

\bibitem{daCunha:2015ana}
B.~Carneiro~da Cunha and F.~Novaes, {\it {Kerr Scattering Coefficients via
  Isomonodromy}},  {\em JHEP} {\bf 11} (2015) 144,
  [\href{http://arxiv.org/abs/1506.06588}{{\tt arXiv:1506.06588}}].

\bibitem{Barragan-Amado:2018pxh}
J.~Barrag\'an-Amado, B.~Carneiro~da~Cunha, and E.~Pallante, {\it {Scalar
  quasinormal modes of Kerr-AdS$\mathbf{_5}$}},
  \href{http://arxiv.org/abs/1812.08921}{{\tt arXiv:1812.08921}}.

\bibitem{Novaes:2018fry}
F.~Novaes, C.~Marinho, M.~Lencs{\'e}s, and M.~Casals, {\it {Kerr-de Sitter
  Quasinormal Modes via Accessory Parameter Expansion}},
  \href{http://arxiv.org/abs/1811.11912}{{\tt arXiv:1811.11912}}.

\bibitem{Novaes:2014lha}
F.~Novaes and B.~Carneiro~da Cunha, {\it {Isomonodromy, Painlev{\'e}
  transcendents and scattering off of black holes}},  {\em JHEP} {\bf 1407}
  (2014) 132, [\href{http://arxiv.org/abs/1404.5188}{{\tt arXiv:1404.5188}}].

\bibitem{Alday:2009aq}
L.~F. Alday, D.~Gaiotto, and Y.~Tachikawa, {\it {Liouville Correlation
  Functions from Four-dimensional Gauge Theories}},  {\em Lett.Math.Phys.} {\bf
  91} (2010) 167--197, [\href{http://arxiv.org/abs/0906.3219}{{\tt
  arXiv:0906.3219}}].

\bibitem{Nekrasov:2002qd}
N.~A. Nekrasov, {\it {Seiberg-Witten prepotential from instanton counting}},
  {\em Adv. Theor. Math. Phys.} {\bf 7} (2003), no.~5 831--864,
  [\href{http://arxiv.org/abs/hep-th/0206161}{{\tt hep-th/0206161}}].

\bibitem{Alba:2010qc}
V.~A. Alba, V.~A. Fateev, A.~V. Litvinov, and G.~M. Tarnopolskiy, {\it {On
  combinatorial expansion of the conformal blocks arising from AGT
  conjecture}},  {\em Lett.Math.Phys.} {\bf 98} (2011) 33--64,
  [\href{http://arxiv.org/abs/1012.1312}{{\tt arXiv:1012.1312}}].

\bibitem{Jimbo:1981aa}
M.~Jimbo, T.~Miwa, and A.~K. Ueno, {\it {Monodromy Preserving Deformation of
  Linear Ordinary Differential Equations With Rational Coefficients, I}},  {\em
  Physica} {\bf D2} (1981) 306--352.

\bibitem{Jimbo:1981ab}
M.~Jimbo and T.~Miwa, {\it {Monodromy Preserving Deformation of Linear Ordinary
  Differential Equations with Rational Coefficients, II}},  {\em Physica} {\bf
  D2} (1981) 407--448.

\bibitem{Jimbo:1981ac}
M.~Jimbo and T.~Miwa, {\it {Monodromy Preserving Deformation of Linear Ordinary
  Differential Equations with Rational Coefficients, III}},  {\em Physica} {\bf
  D4} (1981) 26--46.

\bibitem{Miwa:1981aa}
T.~Miwa, {\it {Painlev{\'e} property of monodromy preserving deformation
  equations and the analyticity of $\tau$ functions}},  {\em Publications of
  the Research Institute for Mathematical Sciences} {\bf 17} (1981), no.~2
  703--721.

\bibitem{Gamayun:2013auu}
O.~Gamayun, N.~Iorgov, and O.~Lisovyy, {\it {How instanton combinatorics solves
  Painlev{\'e} VI, V and IIIs}},  {\em J.Phys.} {\bf A46} (Feb., 2013) 335203,
  [\href{http://arxiv.org/abs/1302.1832}{{\tt arXiv:1302.1832}}].

\bibitem{Iorgov:2014vla}
N.~Iorgov, O.~Lisovyy, and J.~Teschner, {\it {Isomonodromic tau-functions from
  Liouville conformal blocks}},  \href{http://arxiv.org/abs/1401.6104}{{\tt
  arXiv:1401.6104}}.

\bibitem{Gavrylenko:2016zlf}
P.~Gavrylenko and O.~Lisovyy, {\it {Fredholm Determinant and Nekrasov Sum
  Representations of Isomonodromic Tau Functions}},  {\em Commun. Math. Phys.}
  {\bf 363} (2018) 1--58, [\href{http://arxiv.org/abs/1608.00958}{{\tt
  arXiv:1608.00958}}].

\bibitem{Hijano:2015zsa}
E.~Hijano, P.~Kraus, E.~Perlmutter, and R.~Snively, {\it {Witten Diagrams
  Revisited: The AdS Geometry of Conformal Blocks}},
  \href{http://arxiv.org/abs/1508.00501}{{\tt arXiv:1508.00501}}.

\bibitem{Amado:2017kao}
J.~B. Amado, B.~Carneiro~da Cunha, and E.~Pallante, {\it {On the Kerr-AdS/CFT
  correspondence}},  {\em JHEP} {\bf 08} (2017) 094,
  [\href{http://arxiv.org/abs/1702.01016}{{\tt arXiv:1702.01016}}].

\bibitem{Nekrasov:2011bc}
N.~Nekrasov, A.~Rosly, and S.~Shatashvili, {\it {Darboux coordinates, Yang-Yang
  functional, and gauge theory}},  {\em Nucl.Phys.Proc.Suppl.} {\bf 216} (Mar.,
  2011) 69--93, [\href{http://arxiv.org/abs/1103.3919}{{\tt arXiv:1103.3919}}].

\bibitem{Litvinov:2013sxa}
A.~Litvinov, S.~Lukyanov, N.~Nekrasov, and A.~Zamolodchikov, {\it {Classical
  Conformal Blocks and Painleve VI}},
  \href{http://arxiv.org/abs/1309.4700}{{\tt arXiv:1309.4700}}.

\bibitem{Gaiotto:2009ma}
D.~Gaiotto, {\it {Asymptotically free $\mathcal{N} = 2$ theories and irregular
  conformal blocks}},  {\em J. Phys. Conf. Ser.} {\bf 462} (2013), no.~1
  012014, [\href{http://arxiv.org/abs/0908.0307}{{\tt arXiv:0908.0307}}].

\bibitem{Nagoya:2010yb}
H.~Nagoya and J.~Sun, {\it {Confluent primary fields in the conformal field
  theory}},  {\em J. Phys.} {\bf A43} (2010) 465203,
  [\href{http://arxiv.org/abs/1002.2598}{{\tt arXiv:1002.2598}}].

\bibitem{Nagoya:2015cja}
H.~Nagoya, {\it {Irregular conformal blocks, with an application to the fifth
  and fourth Painlev equations}},  {\em J. Math. Phys.} {\bf 56} (2015), no.~12
  123505, [\href{http://arxiv.org/abs/1505.02398}{{\tt arXiv:1505.02398}}].

\bibitem{Jimbo:1982aa}
M.~Jimbo, {\it {Monodromy Problem and the boundary condition for some
  Painlev{\'e} equations}},  {\em Publ. Res. Inst. Math. Sci.} {\bf 18} (1982)
  1137--1161.

\bibitem{Andreev:2000aa}
F.~V. Andreev and A.~V. Kitaev, {\it {Connection formulas for asymptotics of
  the fifth Painlev{\'e} transcendent on the real axis}},  {\em Nonlinearity}
  {\bf 13} (2000), no.~5 1801--1840.

\bibitem{Okamoto:1987aa}
K.~Okamoto, {\it Studies on the painlev\'e equations ii. fifth painlev\'e
  equation p$_v$},  {\em Japanese journal of mathematics. New series} {\bf 13}
  (1987), no.~1 47--76.

\bibitem{Lisovyy:2018mnj}
O.~Lisovyy, H.~Nagoya, and J.~Roussillon, {\it {Irregular conformal blocks and
  connection formulae for Painlev V functions}},  {\em J. Math. Phys.} {\bf 59}
  (2018), no.~9 091409, [\href{http://arxiv.org/abs/1806.08344}{{\tt
  arXiv:1806.08344}}].

\bibitem{Zamolodchikov:1987aa}
A.~B. Zamolodchikov, {\it {Conformal Symmetry in Two-Dimensions: Recursion
  Representation of Conformal Block}},  {\em Teor. Mat. Fiz.} {\bf 73} (1987),
  no.~1 103. Theor. Math. Phys., 53 (1987) 1088.

\bibitem{Seidel:1988ue}
E.~Seidel, {\it {A Comment on the Eigenvalues of Spin Weighted Spheroidal
  Functions}},  {\em Class. Quant. Grav.} {\bf 6} (1989) 1057.

\bibitem{Gamayun:2012ma}
O.~Gamayun, N.~Iorgov, and O.~Lisovyy, {\it {Conformal field theory of
  Painlev\'e VI}},  {\em JHEP} {\bf 1210} (July, 2012) 038,
  [\href{http://arxiv.org/abs/1207.0787}{{\tt arXiv:1207.0787}}].

\bibitem{Ribault:2014hia}
S.~Ribault, {\it {Conformal field theory on the plane}},
  \href{http://arxiv.org/abs/1406.4290}{{\tt arXiv:1406.4290}}.

\bibitem{Leaver:1985ax}
E.~Leaver, {\it {An Analytic representation for the quasi normal modes of Kerr
  black holes}},  {\em Proc.Roy.Soc.Lond.} {\bf A402} (1985) 285--298.

\bibitem{Ronveaux:1995}
A.~Ronveaux and F.~Arscott, {\em Heun's differential equations}.
\newblock Oxford University Press, 1995.

\bibitem{Dubrovin:2018aa}
B.~Dubrovin and A.~Kapaev, {\it {A Riemann-Hilbert approach to the Heun
  equation}},  \href{http://arxiv.org/abs/1809.02311}{{\tt arXiv:1809.02311}}.

\bibitem{Dunne:2019aqp}
G.~V. Dunne, {\it {Resurgence, Painleve Equations and Conformal Blocks}},
  \href{http://arxiv.org/abs/1901.02076}{{\tt arXiv:1901.02076}}.

\end{thebibliography}

\providecommand{\href}[2]{#2}\begingroup\raggedright\endgroup

\end{document}